\documentclass[twoside,twocolumn,9pt]{article}
\pdfoutput=1
\usepackage{extsizes}
\usepackage[super,sort&compress,comma]{natbib} 
\usepackage[version=3]{mhchem}
\usepackage[left=1.5cm, right=1.5cm, top=1.785cm, bottom=2.0cm]{geometry}
\usepackage{balance}
\usepackage{times,mathptmx}
\usepackage{sectsty}
\usepackage{notes2bib}
\usepackage{graphicx}
\graphicspath{{./figures/}} 
\usepackage{lastpage}
\usepackage[format=plain,justification=justified,singlelinecheck=false,font={stretch=1.125,small,sf},labelfont=bf,labelsep=space]{caption}
\usepackage{float}
\usepackage{fancyhdr}
\usepackage{fnpos}
\usepackage[english]{babel}
\addto{\captionsenglish}{%
  \renewcommand{\refname}{Notes and references}
}
\usepackage{array}
\usepackage{droidsans}
\usepackage{charter}
\usepackage[T1]{fontenc}
\usepackage[usenames,dvipsnames]{xcolor}
\usepackage{setspace}
\usepackage[compact]{titlesec}
\usepackage{hyperref}

\usepackage{siunitx}

\definecolor{cream}{RGB}{222,217,201}

\begin{document}

\pagestyle{fancy}
\thispagestyle{plain}
\fancypagestyle{plain}{

\fancyhead[C]{\includegraphics[width=18.5cm]{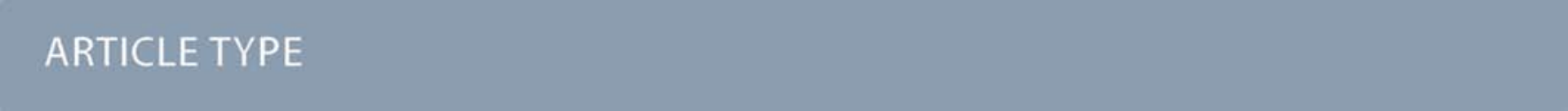}}
\fancyhead[L]{\hspace{0cm}\vspace{1.5cm}\includegraphics[height=30pt]{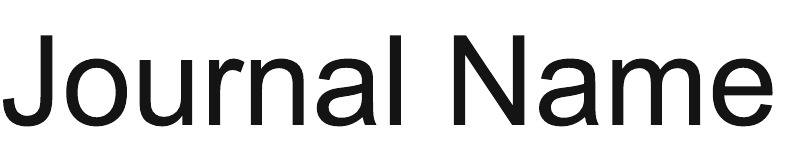}}
\fancyhead[R]{\hspace{0cm}\vspace{1.7cm}\includegraphics[height=55pt]{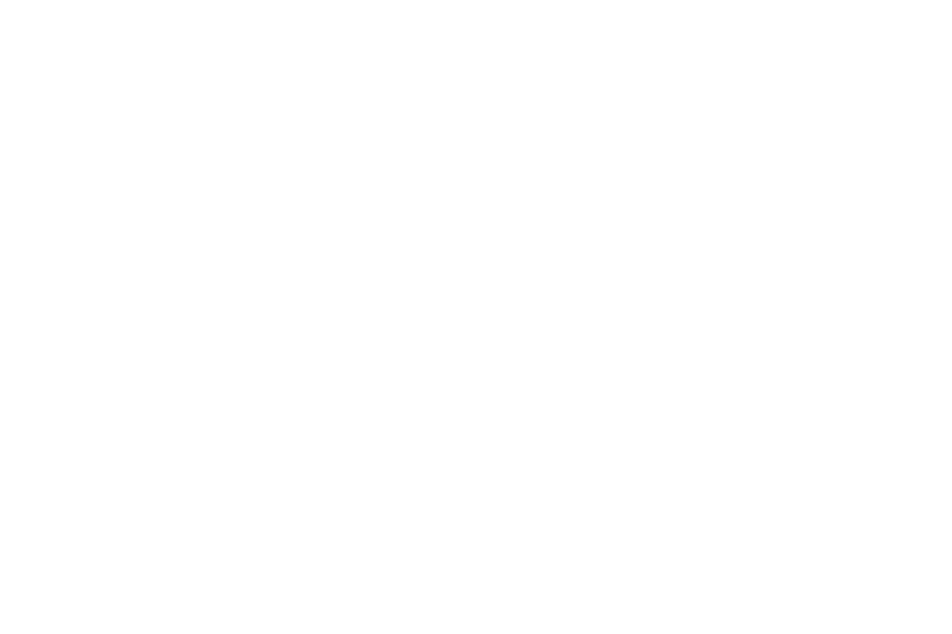}}
\renewcommand{\headrulewidth}{0pt}
}

\makeFNbottom
\makeatletter
\renewcommand\LARGE{\@setfontsize\LARGE{15pt}{17}}
\renewcommand\Large{\@setfontsize\Large{12pt}{14}}
\renewcommand\large{\@setfontsize\large{10pt}{12}}
\renewcommand\footnotesize{\@setfontsize\footnotesize{7pt}{10}}
\makeatother

\renewcommand{\thefootnote}{\fnsymbol{footnote}}
\renewcommand\footnoterule{\vspace*{1pt}%
\color{cream}\hrule width 3.5in height 0.4pt \color{black}\vspace*{5pt}} 
\setcounter{secnumdepth}{5}

\makeatletter 
\renewcommand\@biblabel[1]{#1}            
\renewcommand\@makefntext[1]%
{\noindent\makebox[0pt][r]{\@thefnmark\,}#1}
\makeatother 
\renewcommand{\figurename}{\small{Fig.}~}
\sectionfont{\sffamily\Large}
\subsectionfont{\normalsize}
\subsubsectionfont{\bf}
\setstretch{1.125} 
\setlength{\skip\footins}{0.8cm}
\setlength{\footnotesep}{0.25cm}
\setlength{\jot}{10pt}
\titlespacing*{\section}{0pt}{4pt}{4pt}
\titlespacing*{\subsection}{0pt}{15pt}{1pt}

\fancyfoot{}
\fancyfoot[LO,RE]{\vspace{-7.1pt}\includegraphics[height=9pt]{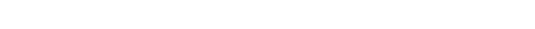}}
\fancyfoot[CO]{\vspace{-7.1pt}\hspace{13.2cm}\includegraphics{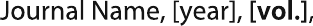}}
\fancyfoot[CE]{\vspace{-7.2pt}\hspace{-14.2cm}\includegraphics{RF}}
\fancyfoot[RO]{\footnotesize{\sffamily{1--\pageref{LastPage} ~\textbar  \hspace{2pt}\thepage}}}
\fancyfoot[LE]{\footnotesize{\sffamily{\thepage~\textbar\hspace{3.45cm} 1--\pageref{LastPage}}}}
\fancyhead{}
\renewcommand{\headrulewidth}{0pt} 
\renewcommand{\footrulewidth}{0pt}
\setlength{\arrayrulewidth}{1pt}
\setlength{\columnsep}{6.5mm}
\setlength\bibsep{1pt}

\makeatletter 
\newlength{\figrulesep} 
\setlength{\figrulesep}{0.5\textfloatsep} 

\newcommand{\topfigrule}{\vspace*{-1pt}%
\noindent{\color{cream}\rule[-\figrulesep]{\columnwidth}{1.5pt}} }

\newcommand{\botfigrule}{\vspace*{-2pt}%
\noindent{\color{cream}\rule[\figrulesep]{\columnwidth}{1.5pt}} }

\newcommand{\dblfigrule}{\vspace*{-1pt}%
\noindent{\color{cream}\rule[-\figrulesep]{\textwidth}{1.5pt}} }

\makeatother

\twocolumn[
  \begin{@twocolumnfalse}
\vspace{3cm}
\sffamily
\begin{tabular}{m{4.5cm} p{13.5cm} }

\includegraphics{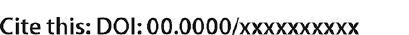} & \noindent\LARGE{\textbf{Active mixtures in a narrow channel: Motility diversity changes cluster sizes}} \\
\vspace{0.3cm} & \vspace{0.3cm} \\

 & \noindent\large{Pablo de Castro$^{\ast a}$, Saulo Diles$^{b}$, Rodrigo Soto$^{a}$ and Peter Sollich$^{c, d}$} \\

\includegraphics{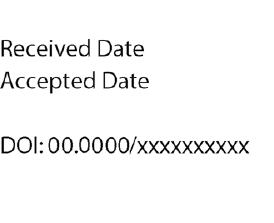} & \noindent\normalsize{The persistent motion of bacteria produces clusters with a stationary cluster size distribution (CSD). Here we develop a minimal model for bacteria in a narrow channel to assess the relative importance of motility diversity (i.e.\ polydispersity in motility parameters) and confinement. A mixture of run-and-tumble particles with a distribution of tumbling rates (denoted generically by $\alpha$) is considered on a 1D lattice. Particles facing each other cross at constant rate, rendering the lattice quasi-1D. To isolate the role of diversity, the global average $\alpha$ stays fixed. For a binary mixture with no particle crossing, the average cluster size ($L_\text{c}$) increases with the diversity as lower-$\alpha$ particles trap higher-$\alpha$ ones for longer. At finite crossing rate, particles escape from the clusters sooner, making $L_\text{c}$ smaller and the diversity less important, even though crossing can enhance demixing of particle types between the cluster and gas phases. If the crossing rate is increased further, the clusters become controlled by particle crossing. We also consider an experiment-based continuous distribution of tumbling rates, revealing similar physics. Using parameters fitted from experiments with \textit{Escherichia coli} bacteria, we predict that the error in estimating $L_\text{c}$ without accounting for polydispersity is around $60\%$. We discuss how to find a binary system with the same CSD as the fully polydisperse mixture. An effective theory is developed and shown to give accurate expressions for the CSD, the effective $\alpha$, and the average fraction of mobile particles. We give reasons why our qualitative results are expected to be valid for other active matter models and discuss the changes that would result from polydispersity in the active speed rather than in the tumbling rate.} \\

\end{tabular}

 \end{@twocolumnfalse} \vspace{0.6cm}

  ]

\renewcommand*\rmdefault{bch}\normalfont\upshape
\rmfamily
\section*{}
\vspace{-1cm}


\footnotetext{\textit{$^{a}$Departamento de F\'{i}sica, Facultad de Ciencias F\'{i}sicas y Matem\'{a}ticas, Universidad de Chile, Avenida Blanco Encalada 2008, Santiago, Chile. Email: pdecastro@ing.uchile.cl}}
\footnotetext{\textit{$^{b}$Faculdade de F\'isica, Universidade Federal do Par\'a, Campus Salin\'opolis, Rua Raimundo Santana Cruz S/N, 68721-000, Salin\'opolis, Par\'a, Brazil}}
\footnotetext{\textit{$^{c}$Disordered Systems Group, Department of Mathematics, King's College London, London, United Kingdom}}
\footnotetext{\textit{$^{d}$Institut f\"ur Theoretische Physik, Georg-August-Universit\"at G\"ottingen, 37077 G\"ottingen, Germany}}



\section{Introduction}
In a process called motility-induced phase separation (MIPS) a homogeneous fluid of self-propelled particles can separate into dense and dilute regions even without attraction.\cite{cates2015motility,soto2014run,sepulveda2016coarsening} For instance, consider a persistent self-propelled particle changing its propulsion direction stochastically. Due to collisions with boundaries or other particles, the particle can get stuck for a long time. This allows for more self-propelled particles to join in and thus block the first particle's way out. If the propulsion direction is sufficiently persistent, this trapping process continues until large clusters are formed,\cite{ginot2018aggregation,redner2016classical} i.e.\ the system undergoes phase separation.

For bacteria, MIPS-like mechanisms can induce the formation of so-called biofilms, making a colony more resilient against antibiotics.\cite{grobas2020swarming} Quasi-two-dimensional experiments have revealed that gliding bacteria (\textit{Myxococcus xanthus} mutants with suppressed biochemical signalling) produce, in their steady state, an exponentially decaying cluster size distribution (CSD) modulated by a power law.\cite{peruani2012collective} Experiments and simulations with active colloids show the same CSD behaviour.\cite{buttinoni2013dynamical,levis2014clustering} In both systems, there is also a high-density transition where a strong peak at high cluster sizes appears in the CSD, in which case the terminology \emph{phase separation} is even more appropriate.

One-dimensional (1D) models are currently a subject of increasing interest.\cite{dandekar2020hard,vanhille2019collective,caprini2020time} In many cases, such systems produce, at least approximately, a purely exponential CSD, as shown across models of run-and-tumble bacteria, active Brownian particles, and active Ornstein-Uhlenbeck particles, both on-lattice and off-lattice.\cite{soto2014run,D0SM00687D} For a simple on-lattice model, it has been demonstrated that no high-density transition exists,\cite{slowman2016jamming} but other lattice models do phase separate more conventionally, e.g.\ by having a typical cluster size instead of having only an exponential distribution across all sizes. This is the case when particles can partially overlap.\cite{sepulveda2016coarsening} For models where the particles are appreciably capable of collectively moving and thus merging clusters by pushing each other,\cite{illien2020speed, barberis2019phase} the CSD is no longer a simple exponential.~\cite{vanhille2019collective} No general theory of 1D active systems exists. Understanding the CSD and its associated average cluster size, which sets the exponential decay scale, is already an enlightening task even in the purely exponential cases.

However, two important features to the CSD of 1D or quasi-1D systems have been overlooked so far, both in analytical approaches and in simulations. First, in any typical bacteria population, there is actually a broad dispersion of motility parameters, i.e.\ the bacteria are not identical swimmers. This is true not only for systems that mix bacteria from different biological groups \cite{whitman1998prokaryotes} but also for bacteria from the same biological species or strain.\cite{tu2005white,andrea2020} For run-and-tumble bacteria, one can then consider that the tumbling rate is not the same for all particles. Instead, the system has a distribution of tumbling rates \cite{andrea2020}---even if commonly only the global average tumbling rate is considered. This motility diversity ingredient is expected to be remarkably relevant. For passive fluids, polydispersity in size or interaction strength crucially alters phase separation.\cite{PabloPeter1,warren1999phase,PabloPeter2,PabloPeter3,decastro2019} Even in the case of active fluids themselves, polydispersity in one or another motility attribute has been shown to produce new phase coexistence properties, as well as novel pattern formation phenomena.\cite{stenhammar2015activity,kolb2020active,hoell2019multi,wittkowski2017nonequilibrium,takatori2015theory,grosberg2015nonequilibrium,curatolo2020cooperative,wang2020phase,van2020predicting}

Second, models of self-propelled particles in one dimension have the potential to partially mimic the behaviour of active particles inside a narrow channel.\cite{quelas2016swimming} Most soil bacteria live in pores of size $\SI{6}{\micro\meter}$ and smaller,\cite{ranjard2001quantitative,mannik2009bacterial} with many of them acting as biofertilizers.\cite{fuentes2005bacterial} In channel geometries, bacteria are expected to eventually swap their positions along the narrow channel provided that its width is larger than the bacterial head. Note that the width does not need to be larger than \emph{two} bacterium heads, because these microorganisms have flexible bodies.
Since bacteria colonies are composed of effectively distinct motility types, taking into account this position crossing feature becomes even more relevant, as otherwise the spatial sequence of particle types would remain fixed at all times. To the best of our knowledge, this effect has not been considered in previous cluster size calculations.

In this work we fill these two gaps at once, both analytically and numerically. We focus on a minimal model of bacteria in a narrow channel that we use to assess the relative importance of motility diversity (i.e.\ polydispersity in the motility parameters) and confinement. 

\begin{figure}[!h]
	\centering
	\includegraphics[width=\columnwidth]{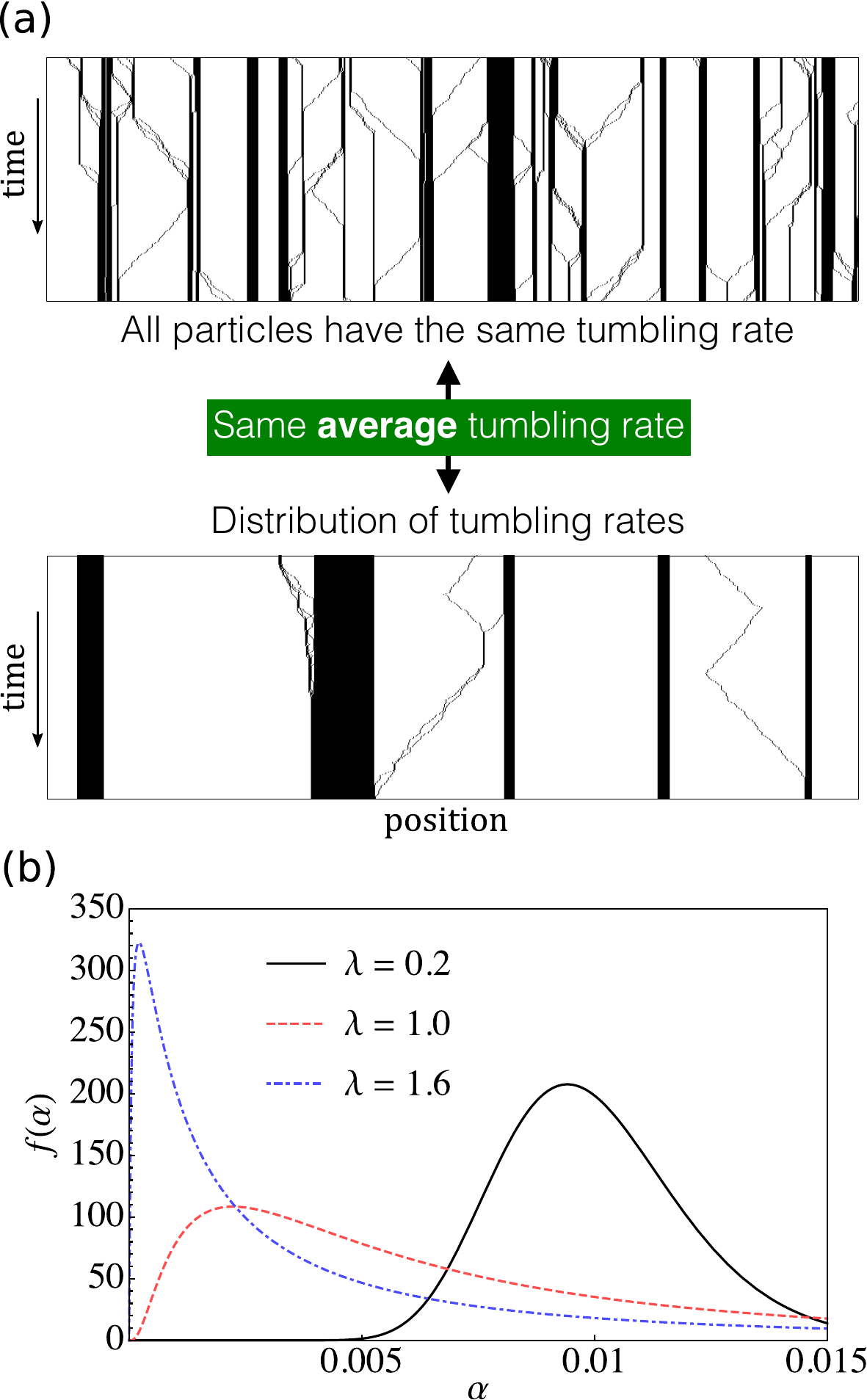}
	\caption{(a) Steady state of a monodisperse system (top) of run-and-tumble particles on a 1D lattice compared against its mixture counterpart with polydispersity in the tumbling rates (bottom); particles cannot cross ($p_\text{cross}=0$). Only $1000$ of $N=2000$ simulated lattice sites are shown. Considering the entire simulated system, the global concentration and the average tumbling rate in both cases are the same: $\phi=0.2$ and $\langle\alpha\rangle\approx0.01$, respectively. In the mixture case (bottom) the polydisperse distribution parameter [see eqn~\eqref{polydist}] is $\lambda=2$. Time flows downwards along the vertical axis for $300$ time steps. Positions are on the horizontal axis. Mobile particles move with speed $\approx1$. Vacancies are in white and particles in black. (b) Polydisperse distribution of tumbling rates, eqn~\eqref{polydist}, for three different distribution parameters $\lambda$ at fixed average tumbling rate $\langle\alpha\rangle=\alpha_0\exp{(\lambda^2/2)}=0.01$.} 
	\label{Fig1}
\end{figure}

Our goal is to identify major physical mechanisms by using a simple active matter model.
To do so, we consider a mixture of run-and-tumble particles on a 1D discrete lattice interacting only via excluded volume, i.e.\ they do not move each other by pushing. The distinct particle types are characterized by their own tumbling rates. To extract the physics more clearly, we first consider a binary system. Next, we discuss the case of an experiment-based continuous distribution of tumbling rates. Moreover, in the binary case we allow particles moving directly towards each other to cross at a constant rate, therefore mimicking the effects of a ``soft'' confinement where particles can swap their positions along the quasi-1D channel. We observe a phenomenon of clustering amplification induced solely by tumbling-rate diversity (see Fig.~\ref{Fig1}a). On the other hand, by relaxing the confinement, the clusters are suppressed and motility diversity becomes less important. One can anticipate that, to some extent, this kind of confinement should produce the same general behaviour as in a three-dimensional off-lattice model of particles in a laterally steep parabolic potential.\cite{ribeiro2016active} In fact, our qualitative results are expected to be valid for other active matter models, including off-lattice ones, as discussed in detail below.

This paper is organized as follows. In Section~\ref{model} our discrete lattice model is presented. Sections \ref{simulations} and \ref{theory} bring our main numerical and analytical results, respectively, for each model variant: (i) binary mixture, (ii) binary mixture with particle crossing, and (iii) fully polydisperse mixture. In Section~\ref{equivalence} we discuss how to obtain approximately equal CSDs between two mixtures with different numbers of particle types. Section~\ref{conc} gives our conclusions.

\section{Model}
\label{model}
We start by reviewing the run-and-tumble model presented in ref.~\citenum{soto2014run}, where all particles are identical. Consider a 1D discrete lattice with $N$ sites and periodic boundary conditions. The maximum occupancy per site is one. Each particle has a propulsion director, which can be left or right. The total number of particles is $M = \phi N$, where $\phi$ is the dimensionless global particle concentration. In each time step, $M$ individual particle updates are performed. A particle is selected at random and a new director for this particle is chosen at random, with probability $\alpha$. Thus, the probability to have a \emph{different} director is $\alpha/2$. In case the particle changes its director, we say that a tumble event has occurred. Otherwise, the particle preserves its previous director. Next, if the propulsion director points toward a neighbouring empty site, then the particle moves to this new position. A new particle is then chosen. The updates are sequential. Our units are such that the lattice spacing and the time step are fixed to unity. A particle can be chosen more than once in a single time step, and so the speed of the mobile particles fluctuates around unity. The initial positions are random and mutually excluding. Each particle is also given an initial random director. A cluster is defined as a contiguous group of occupied sites. There are no velocity alignment mechanisms such as in Vicsek-like models of flocking.\cite{vicsek1995novel}

The above model is monodisperse as all particles have the same motility properties, i.e.\ they all have identical swim speed and tumbling rate. Here we consider a more realistic model where the particles have distinct tumbling rates, while their swim speeds continue to be monodisperse. To motivate this, we first introduce the experiment-based case of a continuous, fully polydisperse distribution of tumbling rates. For both the simulations in Section~\ref{simulations} and the theory in Section~\ref{theory}, we then start with the results for the simpler case of a binary mixture before moving on to the fully polydisperse system.

In bacteria like \textit{Escherichia coli}, the tumbling is triggered by a reversion in the rotation (from counter-clockwise, CCW, to clockwise, CW) of one or several flagella. As a result, the flagella bundle disassembles and propulsion thrust is lost.\cite{chen2000torque} By analysing the biochemistry of the molecular motor, Tu and Grinstein proposed that the tumbling process can be described as a two-state activated system, where the free energy barrier for transitions between the CCW and the CW state depends sensitively on the concentration of a protein inside the bacterium called CheY-P, denoted by $[Y]$.\cite{tu2005white} In the Tu--Grinstein model the tumbling rate is $\alpha=\bar\alpha \exp(-G([Y])/k_\mathrm{B}T)$, where $G$ is the free energy barrier and $\bar\alpha$ a constant. Expanding $G$ around the average value $[Y_0]$, they propose
\begin{equation}
	\alpha(X) = \alpha_0 e^{\lambda X}, \label{alphaofX}
\end{equation}
where $X(t)=([Y](t)-[Y_0])/\sigma_Y$ corresponds to the fluctuations in concentration normalized to $\sigma_Y$, the standard deviation of $[Y]$, with $\alpha_0$ absorbing all prefactors. The parameter $\lambda$ is positive \cite{cluzel2000ultrasensitive} and quantifies the system's sensitivity to changes in the protein concentration. Noticing that this protein has a small production rate, one can show that $X(t)$ is well described by an Ornstein-Uhlenbeck process with a long memory time $T$. By tracking several individual \textit{E.~coli} bacteria it has been possible to fit the model parameters to $T=\SI{19.0}{\second}$, $\alpha_0 = \SI{0.216}{\second^{-1}}$, and 
$\lambda = 1.62$.\cite{figueroa20183d}

Here we use the same tumbling model but with the approximation that the time dependence of $X$ can be neglected as the $T$ value for \textit{E.~coli} is significantly larger than $\alpha_0^{-1}$. Thus, the distribution of $X$, which is Gaussian with zero mean and unit variance, remains fixed at all times. As a result, each particle has a constant tumbling rate drawn from the continuous, fully polydisperse lognormal distribution
\begin{equation}
	f(\alpha)=\frac{1}{\sqrt{2 \pi }   \lambda \alpha}\exp{\left(-\frac{\left[\log\left(\alpha/\alpha_0\right)\right]^2}{2 \lambda ^2}\right)}
	\label{polydist}
\end{equation}
where $\alpha_0$ and $\lambda$ are as defined after eqn~\eqref{alphaofX}. This distribution is qualitatively similar to a Schulz-Gamma distribution, which is commonly used to realistically describe passive fluids with polydispersity in molecular size \cite{PabloPeter1} (see Fig.~\ref{Fig1}b). It changes from a single diverging peak at $\alpha=\alpha_0$ for $\lambda\to0$ to a single diverging peak at $\alpha=0$ for $\lambda\to\infty$. To isolate the effects of motility diversity, we keep $\langle \alpha \rangle\equiv\int_{0}^{\infty}\alpha\,f(\alpha)\,\textrm{d}\alpha=\alpha_0\exp{(\lambda^2/2)}$ fixed while the polydispersity degree is changed by varying $\lambda$. The tumbling rates are assigned so that the initial state is randomly homogeneous and well-mixed.

Although $\alpha>1$ is equivalent to $\alpha=1$ in our model (i.e.\  a tumbling attempt takes place at every step), here we explore a parameter region such that $\int_{0}^{1}\alpha f(\alpha)\,\textrm{d}\alpha\approx\int_{0}^{\infty} \alpha f(\alpha)\,\textrm{d}\alpha$, i.e.\ it makes no practical difference that the distribution assigns values of $\alpha>1$ to a few particles. 
Note that the particular functional form or parameters of the tumbling rates distribution do not affect our main qualitative result below, that is, that clusters are on average amplified when tumbling-rate diversity is increased.

For the binary mixture, we consider half the particles with tumbling rate $\alpha_A=\alpha_0(1+\delta)$ and the other half with tumbling rate $\alpha_B=\alpha_0(1-\delta)$. Again, as we change the degree of motility diversity, $\delta$, the average tumbling rate $\langle \alpha \rangle=\alpha_0$ remains fixed. For simplicity, we do not consider mixing proportions other than $50$-$50\%$ as the fully polydisperse distribution used here already covers a more general case.

We then incorporate an extra ingredient in order to mimic the effects of a narrow channel whose width allows for neighbouring bacteria to exchange positions. Consider a particle after it has potentially tumbled and moved but before a new particle selection has occurred. At this point, we allow the particle to cross its neighbour at a constant rate denoted by $p_{\textrm{cross}}$ if, and only if, their directors point toward each other (see the illustration in the inset of Fig.~\ref{Fig5}b). Of course, what  happens in reality is that the particles directly swap their positions along the channel, without any actual particle crossing. This is similar to some setups for phase-separating passive colloids.\cite{PabloPeter1,PabloPeter2} Out of all possible crossing events at each time step, a fraction equal to $2p_{\textrm{cross}}$ will indeed occur (on average), since both participating particles are equally likely to be chosen. 

Our model relies on the assumption that the channel is sufficiently narrow to keep the bacteria moving mostly along the channel axis. Also, we assume that increasing the channel width corresponds to increasing $p_{\textrm{cross}}$. These two quantities are expected to be connected through some nontrivial but monotonic relationship. Thus, we refrain from controlling the channel width directly and instead take $p_{\textrm{cross}}$ as our free parameter. Effectively, this makes the lattice quasi-1D rather than 1D. Of course at channel widths that are too small, tumbling through flagella rearrangement might become impossible to perform due to excessive confinement. On the other hand, if the channel width is too large (high $p_{\textrm{cross}}$), then the particles would essentially be invisible to each other, assuming that biochemical signalling has at most a subdominant effect as in the experimental setup of ref.~\citenum{peruani2012collective}. We will therefore assume that the width takes intermediate values, such that there is enough space for tumbling to take place but with $p_\textrm{cross}$ remaining smaller than one.
As we shall see, considering $p_{\textrm{cross}}\neq0$ affects clustering even in the monodisperse case, i.e.\ regardless of motility diversity. Nevertheless, we focus below on the relative importance of tumbling rate diversity and nonzero $p_{\textrm{cross}}$.

\section{Simulations}
\label{simulations}
Our 1D numerical results were obtained in the stationary regime with $N=2000$ sites and periodic boundary conditions. To visualize the dynamics within the steady state, we recorded snapshots of the system for the first $300$ successive time steps after $t=10^7$. To calculate the CSD and other similarly averaged quantities, we use $9000$ uncorrelated configurations, within the same simulation, obtained every $10^4$ time steps between $t=10^7$ and $t=10^8$, unless otherwise stated. An alternative approach would be as follows: instead of recording the state of the system at different times within the same simulation, we could run \emph{different} simulations all the way from the initial state until $t=10^7$ and record only their final configurations. We have confirmed that both methods lead to quantitatively equivalent results across all used parameter ranges, but the computational cost of the former approach is significantly lower. To avoid undesirable finite-size effects, we use a value of $N$ in the range where the system size no longer affects the normalized CSD within simulation accuracy.

\subsection{Binary mixture}
\begin{figure}[!h]
	\centering
	\includegraphics[width=\columnwidth]{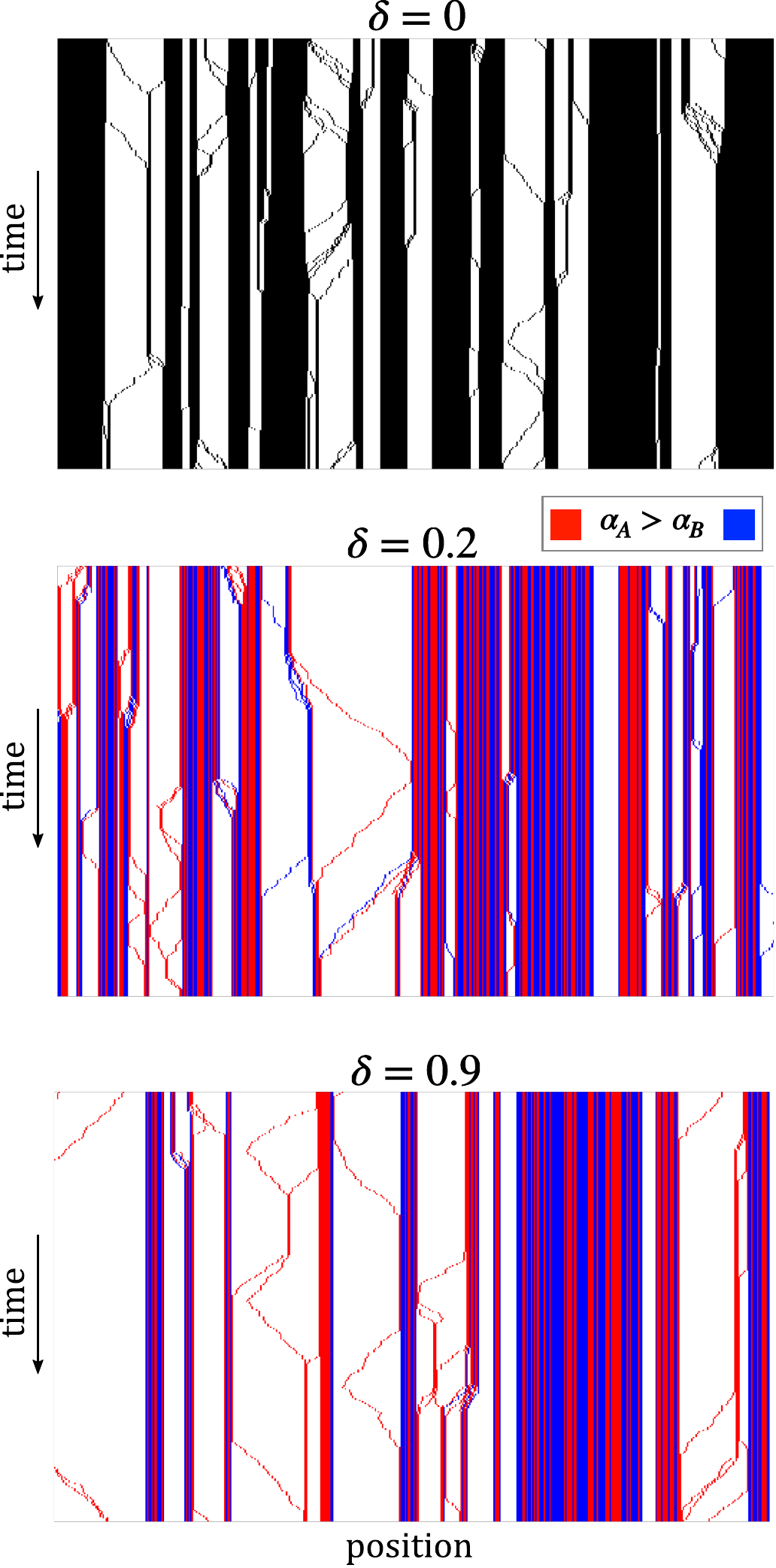}
	\caption{As Fig.~\ref{Fig1}a but for the steady state of a binary system of run-and-tumble particles on a 1D lattice for different bidispersities $\delta$. Only $500$ of $N=2000$ simulated sites are shown. The monodisperse case ($\delta=0$) is shown for comparison. In the bidisperse cases ($\delta>0$), particles with higher (lower) tumbling rate are in red (blue). The global average tumbling rate, concentration and composition of the entire simulated system are $\alpha_0=0.01$, $\phi=0.5$ and $50$-$50\%$, respectively.}
	\label{Fig2}
\end{figure}

We start with $p_{\textrm{cross}}=0$ and the binary mixture model. Fig.~\ref{Fig2} shows 1D snapshots of a section of the simulated system at successive times within the steady state for different bidispersity degrees $\delta$. For $\delta=0$, all particles have the same tumbling rate and therefore the same tendency to form clusters. They run until they either change direction due to tumbling or bump into each other. In the latter case they get blocked and act as seeds for clusters, in which case the tumbling rate has to be sufficiently small, otherwise the particles would change direction before another particle joined and thus trapped them. As time passes from the initial homogeneous state, particles start to trap each other and form clusters, reaching a steady state, which will be characterized by the average cluster size.

\begin{figure}[!h]
	\centering
	\includegraphics[width=\columnwidth]{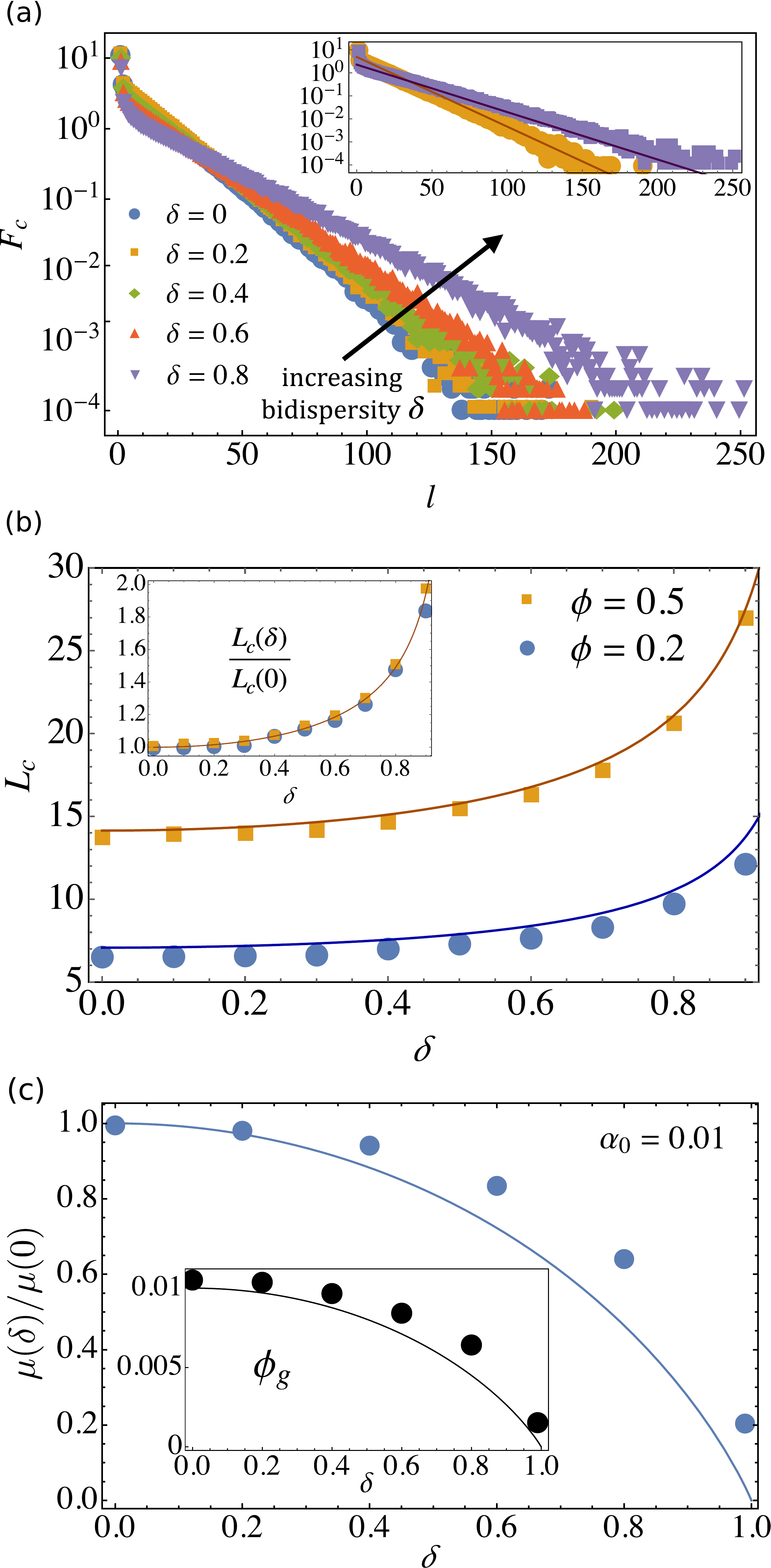}
	\caption{(a) Cluster size distribution (log scale) from simulations for various bidispersities $\delta$, with $\alpha_0=0.01$, $\phi=0.5$, $N=2000$, and $p_\text{cross}=0$. The inset shows the agreement with the corresponding theory (lines) for $\delta=0.2$ and $\delta=0.8$.
		(b) Average cluster size $L_\text{c}$ versus bidispersity $\delta$ at fixed average tumbling rate. The points show the simulation results and the lines are the theoretical predictions for the CSD length scale $l_\text{c}$. The inset shows the ratio to the corresponding monodisperse case, which is independent of $\phi$. Other parameters as in (a). The theoretical expressions are obtained by inserting eqn~\eqref{deltaeffalpha} into eqn~\eqref{monoparam}. (c) Average fraction of mobile particles $\mu$ versus bidispersity $\delta$ normalized by the monodisperse value $\mu(0)=0.014$, from simulation (points) and theory (line). Parameters: $\phi=0.5$ and $\alpha_0=0.01$. The inset shows the gas density $\phi_g$ versus $\delta$ for $\phi=0.2$.}	
	\label{Fig3}
\end{figure}

For $\delta>0$, there are two groups of particles, each with a different tendency to form clusters. Since particle crossing is not allowed for now, the random sequence of particle types remains the same at all times. The composition of each cluster has to rely on both particles types since the probability to have many particles of the same type at successive positions is vanishingly small, except for isolated particles and small clusters, which tend to be dominated by higher-tumbling-rate particles. Once a cluster is formed, the tumbling rates of the particles well inside it are irrelevant for the cluster size. The position of the cluster border is dictated by the tumbling rate of the particle at its boundary, which faces toward its interior by definition. Type $B$ particles (lower tumbling rate) will typically take longer to escape from the cluster---this happens once they become a border particle and perform a tumble---than type $A$ ones. This means that lower-tumbling-rate particles trap higher-tumbling-rate ones for longer times. The cluster sizes are set by the average behaviour between both types, with lower-tumbling-rate particles contributing more. Fig.~\ref{Fig3}a shows the CSD defined as the average number of clusters of size $l$, $F_\text{c}(l)$, for different $\delta$ values. As $\delta$ increases the CSD moves toward bigger clusters, even though the global average tumbling rate $\alpha_0$ remains fixed. This result is related to the fact that each tumbling rate contributes with a length scale proportional to ${\alpha}^{-1/2}$ as we shall see in Section~\ref{theory}. But the exponential shape of the CSD is preserved, with $F_\text{c}(l)\sim\exp{(-l/L_\text{c})}$, where $L_\text{c}$ is the average cluster size.

Fig.~\ref{Fig3}b shows the average cluster size, $L_\text{c}$, which increases with the bidispersity degree $\delta$, for two values of $\phi$. The inset has the ratio between the bidisperse and monodisperse values of $L_\text{c}$. This ratio quantifies the clustering amplification by motility diversity. It does not depend on $\phi$, in agreement with our theory in Section~\ref{theory} below. Similarly, our simulations confirm that the ratio does not depend on $\alpha_0$ either (data not shown). This provides evidence that, in the ``thermodynamic'' limit, $L_\text{c}$ diverges as $\delta\to1$ since one of the particle types becomes infinitely persistent while still corresponding to an infinitely large number of particles.\cite{mandal2020extreme} This is also supported by our theory. Finally, we notice that clusters of size $l=1$ correspond to isolated particles and therefore should, in principle, be considered as part of the gas. However, the basic theoretical result used as a starting point in Section~\ref{theory} \cite{soto2014run} relies on integrating quantities across all positive $l$. Thus, $l=1$ is included in calculating $L_\text{c}$ for a more appropriate comparison. In any case, since at low tumbling rates the gas density is typically small, the contribution to $L_\text{c}$ from isolated particles is likewise very small (typically $<5\%$), as we have confirmed numerically. 

At each time step only a fraction of the particles can move and the others are jammed. The average fraction of mobile particles, denoted by $\mu$, is plotted in Fig.~\ref{Fig3}c. It decreases with $\delta$ since at higher diversity more particles participate in clusters of size $l>1$, as seen in Fig.~\ref{Fig3}a. The same occurs with the gas density $\phi_g$ (see inset of Fig.~\ref{Fig3}c), which is defined as the average particle concentration in the regions occupied only by isolated particles (i.e.\ not occupied by clusters of size $\geq2$).

\subsection{Binary mixture with particle crossing}
\begin{figure}[!h]
	\centering
	\includegraphics[width=\columnwidth]{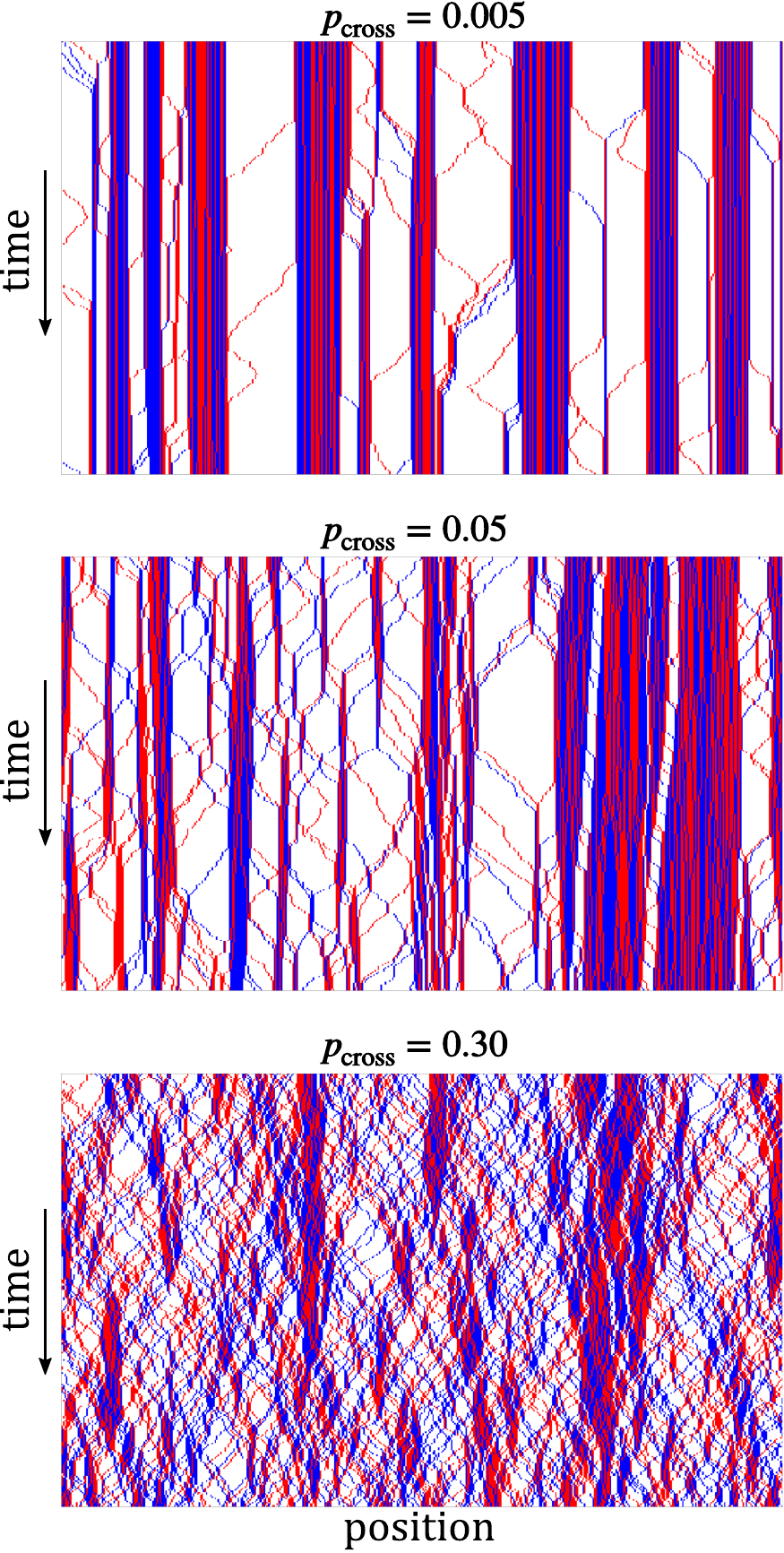}
	\caption{As Fig.~\ref{Fig2} but for fixed $\delta=0.9$ and particle crossing rate $p_{\rm cross}>0$ as indicated.}
	\label{Fig4}
\end{figure}

By turning $p_{\textrm{cross}}$ on, particles can now escape from the clusters more quickly. At each time step, they have a higher chance to escape as now this can occur either by tumbling or by crossing. As we shall see in Section~\ref{theory}, doing so effectively increases the tumbling rate. Fig.~\ref{Fig4} shows snapshots within the steady state for different $p_{\textrm{cross}}$ values. The higher the crossing rate, the more the cluster sizes fluctuate in time: since particles can now cross, cluster formation is suppressed. At high $p_{\textrm{cross}}$, the clusters have been mostly destroyed.

\begin{figure}[!h]
	\centering
	\includegraphics[width=\columnwidth]{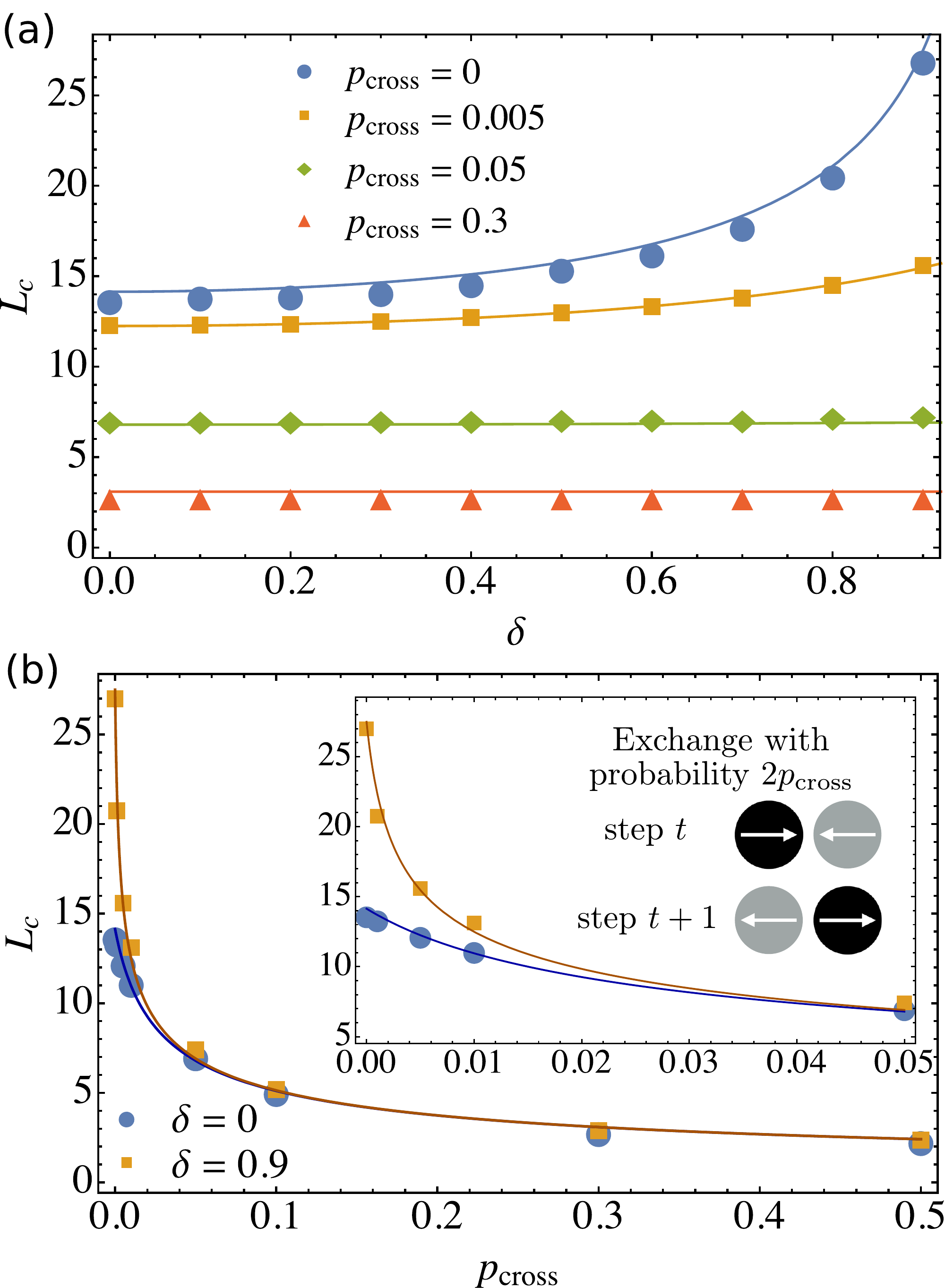}
	\caption{(a) Average cluster size $L_\text{c}$ versus bidispersity $\delta$ at fixed average tumbling rate for various particle crossing rates $p_{\rm cross}$, from simulations (points) and theoretical predictions for the CSD length scale $l_\text{c}$ (lines). (b) Same but for two fixed bidispersities, $\delta=0$ and $\delta=0.9$, as a function of $p_{\rm cross}\in\{0, 0.001, 0.005, 0.01, 0.05, 0.1, 0.3, 0.5\}$. The inset is a zoom-in of the low $p_{\rm cross}$ region where $L_\text{c}$ is significantly affected by $\delta$. Other parameters as in Fig.~\ref{Fig3}a. The theoretical expressions are obtained by inserting eqn~\eqref{eq.alphaeff.deltacross} into eqn~\eqref{monoparam}. The illustration inside the inset in (b) shows a particle crossing event.
	}
	\label{Fig5}
\end{figure}

The above picture is supported by our numerical results for the CSD, which recede toward low cluster sizes while maintaining an approximately purely exponential functional form. Focusing on the average cluster size, Fig.~\ref{Fig5}a shows $L_\text{c}$ versus $\delta$ for various values of $p_{\textrm{cross}}$, whereas Fig.~\ref{Fig5}b shows $L_\text{c}$ versus $p_{\textrm{cross}}$ for fixed values of $\delta$. For high $p_{\textrm{cross}}$, the cluster size dependence on $\delta$ is negligible. This is because, on average, the particles end up leaving the cluster sooner by moving past other particles than they would do by tumbling. Particles with different tumbling rates across the system do not have enough time to perform many tumbles. As a result, it does not matter whether the system is bidisperse or not, hence the effect of $\delta$ becomes negligible in this extreme scenario. The clustering process becomes primarily controlled by crossing events.

It is worth mentioning that, in a real system with a sufficiently high channel width, the clusters should no longer be characterized solely by their linear size. Instead, since two or more bacteria could be located simultaneously at the same point along the channel axis, one should use the number of bacteria participating in a cluster. Any associated additional effects are neglected in our simulations here.

\subsection{Fully polydisperse mixture}
Finally, consider the fully polydisperse distribution of tumbling rates in eqn~\eqref{polydist}. For simplicity, we keep $p_{\textrm{cross}}=0$. Any particle crossing effects are expected to be analogous to those in the binary mixture.

\begin{figure}[!h]
	\centering
	\includegraphics[width=\columnwidth]{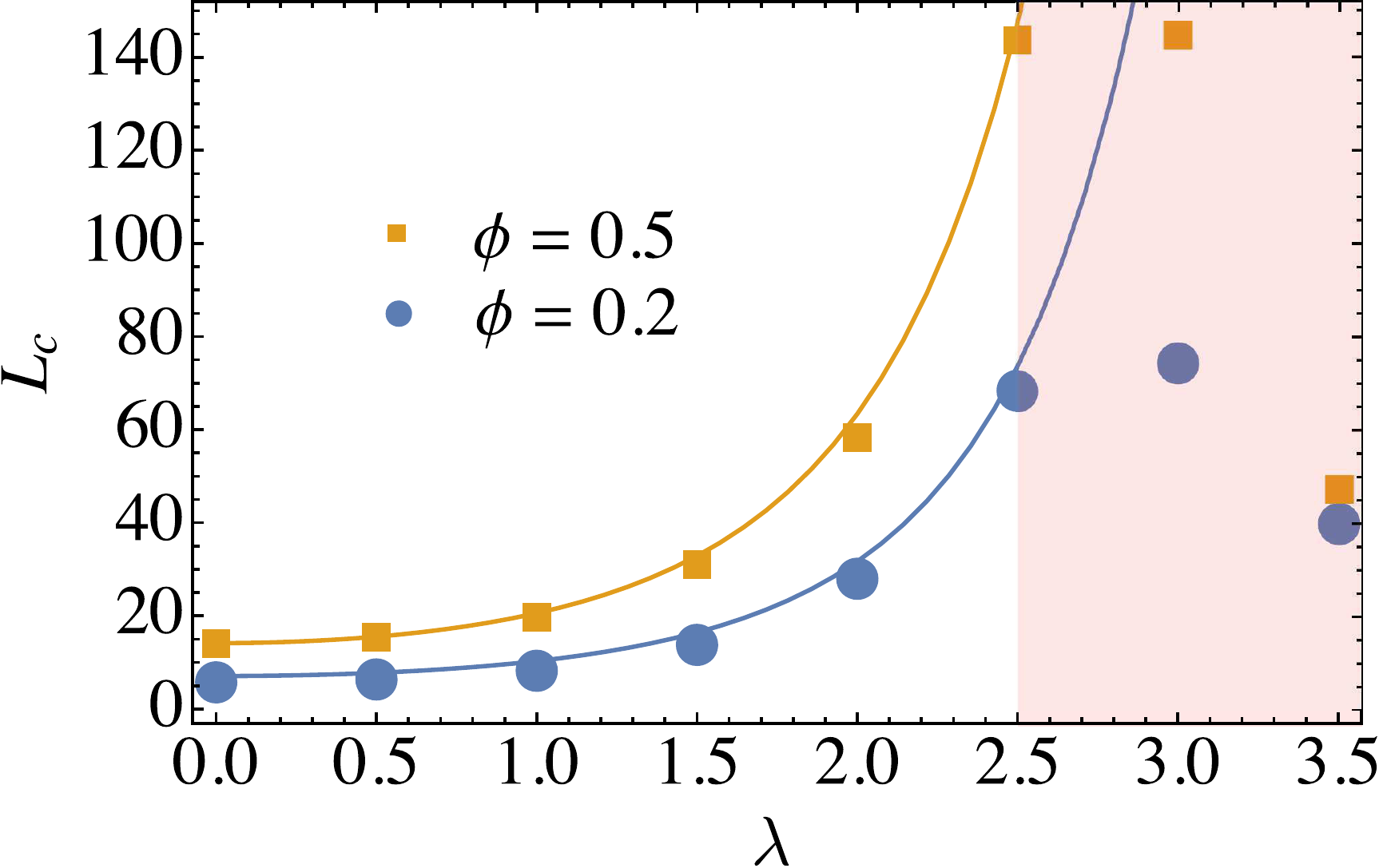}
	\caption{Average cluster size $L_\text{c}$ versus the polydispersity parameter $\lambda$ at fixed average tumbling rate $\langle\alpha\rangle=0.01$ for $\phi=0.2$ and $\phi=0.5$. The points show the simulation results and the lines are the theoretical predictions for the CSD length scale $l_\text{c}$, eqn~\eqref{polylength}. For the simulations with $\lambda>2.5$, the numerical results no longer agree with the theory, as discussed in Section~\ref{theory}.}
	\label{Fig6}
\end{figure}

We find that the resulting cluster sizes behave, in fact, analogously to the binary mixture case. Fig.~\ref{Fig6} shows the average cluster size. The higher the value of $\lambda$, the larger are the clusters, at either fixed $\alpha_0$ or fixed $\langle\alpha\rangle$ (data not shown). The CSD maintains an approximately purely exponential form within the range of validity of the fully polydisperse model, becoming almost horizontal at sufficiently high $\lambda$ (data not shown). Additional details on our numerical results for this case are found in Section~\ref{theory}, where we compare them with our theory.

\section{Theory}
\label{theory}
In ref.~\citenum{soto2014run} it was shown that when ${\alpha\ll\phi}$ cluster-cluster interactions are weak and occur through the uncorrelated emission and absorption of gas
particles; thus, the position of each cluster border undergoes diffusive motion and each cluster evolves independently of the others, except for particle conservation constraints. Here, we work within the same approximation.
In ref.~\citenum{soto2014run} this was used to map the steady-state clustering process onto an equilibrium process for the sizes of independent
pseudoparticles (i.e.\ the clusters), allowing the CSD to be obtained
by maximizing the relevant configurational entropy.
By applying this method to the monodisperse case without particle crossing, it was shown \cite{soto2014run} that the average number of clusters of size $l$ is $F_\text{c}(l) = A_\text{c} e^{-l/l_\text{c}}$ where the prefactor and the length scale are
\begin{equation}
	A_\text{c} \approx \frac{N\alpha_{0}\left(1-\phi\right)}{2},~~~l_\text{c} \approx \sqrt{\frac{2\phi}{\alpha_{0}(1-\phi)}}.
	\label{monoparam}
\end{equation}
A sufficient number of approximate steady-state criteria involving $\alpha_{0}$ and $\phi$ had to be employed to fix the CSD parameters. The explicit formula for $A_\text{c}$ was not given in ref.~\citenum{soto2014run} but follows directly from the analysis there.

To include motility diversity, one could try to develop a generalization to the mixture setting of the aforementioned entropic derivation of the CSD. A simpler approach, which as we will show provides a good approximation, is to consider that the mixture system follows a \emph{monodisperse} CSD with a new, effective tumbling rate $\alpha_{\rm eff}$ that takes into account the tumbling rate distribution. That is, our new CSD is $F_\text{c}(l)=A_\text{c}e^{-l/l_\text{c}}$ with $A_\text{c}$ and $l_\text{c}$ given by eqn~\eqref{monoparam} with $\alpha=\alpha_{\rm eff}$. The length scale $l_\text{c}$ for the mixture is assumed to be approximated by the average of the monodisperse length scales belonging to each particle type, weighted by their particle fraction (this assumption is further clarified in Section~\ref{conc}). For the $50$-$50\%$ binary mixture with ${\alpha_A = \alpha_{0}(1+\delta)}$ and ${\alpha_B = \alpha_{0}(1-\delta)}$, we have
\begin{equation}
	\label{lengthbi}
	l^{\rm{bi}}_\text{c} \approx \sqrt{\frac{\phi}{2\alpha_{0}(1-\phi)}}\left(\frac{1}{\sqrt{1+\delta}}+\frac{1}{\sqrt{1-\delta}}\right)
\end{equation}
where we assumed that the highest tumbling rate is still much smaller than $\phi$. As seen here and in ref.~\citenum{soto2014run}, the average cluster size is well captured by the length scale $l_\text{c}$, i.e.\ $L_\text{c}\approx l_\text{c}$ in both the monodisperse and the mixture cases. eqn~\eqref{lengthbi} predicts that $L_\text{c}$ increases with $\delta$, a result that is corroborated by the simulations.

The effective tumbling rate $\alpha_{\rm eff}$ is the tumbling rate which, in a monodisperse system [eqn~\eqref{monoparam}], gives the same length scale as the mixture system [eqn~(\ref{lengthbi})]. As a result,
\begin{equation}
	\alpha_{\rm eff} = \frac{2 \alpha _0 \left[\left(1-\delta ^2\right)-\left(1-\delta ^2\right)^{3/2}\right]}{\delta ^2}.
	\label{deltaeffalpha}
\end{equation}
In the limit $\delta\to0$, eqn~\eqref{deltaeffalpha} reduces to the monodisperse tumbling rate $\alpha_0$ as expected. When $\delta\to1$, the lower-tumbling-rate particles become infinitely persistent, leading to $\alpha_{\rm eff}=0$ and a diverging $L_\text{c}$ in the ``thermodynamic'' limit of infinite system size.
The effective tumbling rate can now be inserted into the monodisperse expressions to obtain other quantities such as the entire CSD, $F_\text{c}(l)$.

Particle crossing is taken into account through a similar approach. In the monodisperse case, where these effects are already relevant, we find that, to first order in $p_{\rm cross}$, a good approximation for the effective tumbling rate is 
\begin{equation}
	\alpha_{\rm eff}=\alpha_0 + \frac{2p_{\rm cross}}{3}.
	\label{crossingeffalpha}
\end{equation}
To see why, notice that switching on particle crossing enables an additional mechanism for cluster border evaporation, other than tumbling. For clusters of size $l=2$, i.e.\ dimers, only a third of them (at the steady state) will be affected by particle crossing since any other possible configurations, i.e.\ either both particles moving to the right or both to the left, does not contain particles directly facing each other. Neighbouring particles moving away from each other are no longer considered a dimer cluster since, to first order in the tumbling rate, they will separate in the next step. The picture for a pair of particles located at the border of bigger clusters is more complicated; for instance, it depends on higher order terms in the tumbling and particle crossing rates, which are ignored in this derivation. Therefore, we consider that the rate at which a randomly chosen particle evaporates through particle crossing is $p_{\rm cross}/3$. Finally, we add the rate $\alpha_0/2$ at which a particle chooses a different director and evaporates through tumbling, yielding eqn~\eqref{crossingeffalpha}. 

For the bidisperse case, in principle two procedures are possible: (i) one could replace each of the two tumbling rates involved by their effective tumbling rates with respect to particle crossing [eqn~\eqref{crossingeffalpha}] and then calculate the weighted average of the two resulting length scales or (ii) one could first average the length scales with respect to motility diversity and then add the term due to particle crossing. Both procedures have been carried out, giving similar results. However, simulations at high diversity indicate that the former procedure (i) yields a better approximation. We find
\begin{multline}
	l_\text{c}^{\textrm{bi-cross}}=\sqrt{\frac{\phi}{2(1-\phi)}} \\
	\resizebox{.93\hsize}{!}{$\times\left[ 
		\frac{1}{\sqrt{\alpha_0(1+\delta)+\frac{2p_\text{cross}}{3}}}+ \frac{1}{\sqrt{\alpha_0(1-\delta)+\frac{2p_\text{cross}}{3}}}
		\right]$},
\end{multline}
yielding
\begin{align}
	&\alpha_{\rm eff}=\frac{2}{27 \alpha _0^2 \delta ^2} \left[\left(3 \alpha _0+2 p_{\text{cross}}\right){}^2-9 \alpha _0^2 \delta ^2\right] \nonumber \\ 
	&\times\left[3 \alpha _0+2 p_{\text{cross}}-\sqrt{\left(3 \alpha _0+2
		p_{\text{cross}}\right){}^2-9 \alpha _0^2 \delta ^2}\right]. \label{eq.alphaeff.deltacross}
\end{align}
In the limits $p_{\text{cross}}\to0$ and $\delta\to0$, respectively, eqns~\eqref{deltaeffalpha} and \eqref{crossingeffalpha} are recovered. At sufficiently high $p_{\text{cross}}$ and low $\alpha _0$, the dependence of $\alpha_{\rm eff}$ on $\delta$ becomes negligible, as seen in the simulations (see Fig.~\ref{Fig5}). 

The limit $\delta\to1$ no longer diverges: although in this case one of the particle types is infinitely persistent, which would block cluster borders and in an infinite system cause $L_\text{c}$ to diverge, particle crossing still allows particles to escape, therefore keeping the effective tumbling rate nonzero. Even if both particle types were infinitely persistent, e.g.\ if their tumbling systems had been damaged or blocked, we would still have a finite $L_\text{c}$ if ${p_{\text{cross}}\neq0}$.

Despite its simplicity, this effective monodisperse approach leads to satisfactory agreement with numerical simulations (see Figs.~\ref{Fig3} and \ref{Fig5}). At small tumbling rates, since some very large clusters appear, the statistical sampling for those cluster sizes will not necessarily be robust, leading to a slight mismatch between theory and simulations (data not shown).
Still, our theory captures the steady-state cluster sizes with good accuracy. Also, because we have assumed that the tumbling rates satisfy $\alpha\ll\phi$, the case $\phi=0.5$ shows better agreement than for $\phi=0.2$.

Another advantage of the effective monodisperse approach set out above is that one can easily extend it even to the fully polydisperse case. This would be impossible to do by generalizing the entropic derivation due to the number of new equations needed to fix the additional parameters for each new particle type. To do so, we perform the continuum average of the length scale. Assuming now no particle crossing, this gives
\begin{equation}
	l_\text{c}^{\rm{poly}} \approx \int_0^\infty \textrm{d}\alpha\, f(\alpha)\,\sqrt{\frac{2\phi}{\alpha(1-\phi)}}
	\label{polygeneral}
\end{equation}
where $f(\alpha)$ is an arbitrary polydisperse distribution giving the fraction of particles with tumbling rate between $\alpha$ and $\alpha+\textrm{d}\alpha$. For this expression to be valid, there must exist $\alpha=\alpha_{\rm cutoff}\ll\phi$ such that $\int_0^{\alpha_{\rm cutoff}} \textrm{d}\alpha\, f(\alpha)\,l_\text{c}^{\rm{mono}}(\alpha) \approx\int_0^\infty \textrm{d}\alpha\, f(\alpha)\,l_\text{c}^{\rm{mono}}(\alpha)$. Using distribution \eqref{polydist}, we have 
\begin{equation}
	l_\text{c}^{\rm{poly}} \approx \sqrt{\frac{2\phi}{\alpha_0 (1-\phi )}}e^{\frac{\lambda ^2}{8}}=\sqrt{\frac{2\phi}{\langle\alpha\rangle (1-\phi )}}e^{\frac{3\lambda ^2}{8}}.
	\label{polylength}
\end{equation}
Again, the monodisperse ($\lambda\to0$) and infinitely-polydisperse ($\lambda\to\infty$) limits are as expected.  Using eqn~\eqref{monoparam}, the effective tumbling rate is $\alpha_{\rm eff}=\alpha_0   \exp{(-\lambda ^2/4)}$. Within the expected range of validity this is in good agreement with the simulations as shown in Fig.~\ref{Fig6}. For $\lambda>2.5$, the system has many particles with $\alpha>1$ as per the assumed $f(\alpha)$ and even more particles whose tumbling rate does not satisfy $\alpha\ll\phi$. Also, many particles with small tumbling rates arise, making $N$ not large enough to eliminate finite-size effects in those exceptional cases. As a result, the theory no longer agrees with the simulations for that range of $\lambda$.

When the average tumbling rate $\langle\alpha\rangle$ of a polydisperse system is known but $\lambda$ is not, i.e.\ the tumbling rate distribution is unknown, one might be tempted to estimate $L_\text{c}$ via eqn~\eqref{monoparam} by considering that the system is monodisperse and has $\alpha=\langle\alpha\rangle$. The error in doing so can be expressed by comparison with the length scale in eqn~\eqref{polylength} as 
\begin{equation}
	\Delta l_\text{c}\equiv\left(l_\text{c}^{\rm{poly}}-l_\text{c}^{\rm{mono}}\right)/l_\text{c}^{\rm{poly}}= 1-e^{-\frac{3 \lambda ^2}{8}}.
	\label{error}
\end{equation}
This nearly saturates at $100\%$ for $\lambda\approx3.6$. For the \textit{E.~coli} value $\lambda=1.62$,\cite{figueroa20183d} eqn~\eqref{error} means that the relative error associated with not taking polydispersity into account is such that $L_\text{c}$ is $\approx63\%$ smaller in a monodisperse system that has the same average tumbling rate $\langle\alpha\rangle$. This is in quantitative agreement with our simulations. 

Also, the $\alpha_0$ value fitted for \textit{E.~coli} in ref.~\citenum{figueroa20183d} can be used to estimate an absolute value for their $L_\text{c}$. To do so, we first need to express $\alpha_0$ in our model units. Considering that the typical bacterium body size and speed are $\approx\SI{2}{\micro\meter}$ and $\approx\SI{14}{\micro\meter/\second}$,\cite{lovely1975statistical} our units are such that one time step is roughly $\SI{1/7}{\second}$, and thus $\alpha_0 = \SI{0.216}{\second^{-1}}$ corresponds to $\alpha_0\approx0.03$. For $\phi=0.2$, the theoretical estimate via eqn~\eqref{polylength} for the fully polydisperse case with $\lambda=1.62$ is $L_c\approx5.57$. For the monodisperse case ($\lambda=0$) with the same $\langle\alpha\rangle$ and $\phi$ as in the fully polydisperse case, we find $L_c\approx2.11$. Once again, these average cluster size estimates are in units of bacterium body size and take into account clusters of all sizes $l\geq1$, including isolated particles. Since the corresponding average tumbling rate $\langle\alpha\rangle\approx0.11$ is not too small compared with our chosen concentration $\phi=0.2$, the simulation result for the fully polydisperse case is slightly different: $L_c\approx5.15$. If isolated particles are not taken into account, then the monodisperse and fully polydisperse simulation results are $L_c\approx3.65$ and $L_c\approx8.02$, respectively. 

The theory presented above, which finds an effective tumbling rate to match the average cluster size, is also able to give the values of other relevant observables.
The dynamical equilibrium in the monodisperse case is such that the gas density obeys $\phi_g=\alpha_0$ when $\alpha\ll\phi$.\cite{soto2014run} In its generalized version, i.e.\ $\phi_g=\alpha_{\rm eff}$, this equation allows us to calculate the gas density for a mixture with particle crossing. Also, in the monodisperse case, the average mobility fraction $\mu$ is well fitted by the expression $\mu=\alpha(1-\phi)(1 + \phi + 6\phi\alpha)/(\phi+\alpha+6\phi\alpha)$.\cite{soto2014run} For a binary mixture, by inserting $\alpha=\alpha_{\rm eff}(\delta)$, we can readily find an expression for $\mu(\delta)$. The ratio $\mu(\delta)/\mu(0)$ decreases from $1$ to $0$ as we move from $\delta=0$ to $\delta=1$. 
These generalized expressions for $\phi_g$ and $\mu$ are in reasonable agreement with our numerical data as seen in Fig.~\ref{Fig3}c for $p_{\rm cross}=0$. Both quantities are directly related to the number of isolated particles. Because our effective theory is based on matching the \emph{average} cluster size $L_\text{c}$, ignoring other moments of the CSD, it captures $\mu$ and $\phi_g$ less well at high $\delta$. This could be anticipated by noticing that the change of behaviour near $l=1$ in the CSDs is stronger for higher $\delta$, as shown in Fig.~\ref{Fig3}. 

With particle crossing switched on, particle mobility \emph{within clusters} is enabled. This was impossible in the case without crossing, where an internal cluster dynamics existed only in terms of ineffective tumble events but no particle motion. Each crossing event contributes with two particles to the mobility fraction. The dependence of $\mu$ on $p_{\rm cross}$ is strong and becomes dominant, i.e.\ $\mu$ depends only weakly on $\delta$ at high $p_{\rm cross}$. By using eqn~\eqref{eq.alphaeff.deltacross} for the effective tumbling rate $\alpha_{\rm eff}(\delta,p_{\rm cross})$ in the fitted expression for $\mu$, we were able to reasonably reproduce the simulation results for the mobility fraction (data not shown), provided that particle moves due to crossing are not counted. This is expected since our effective theory has no particle crossing, i.e.\ it cannot capture internal cluster crossing events, but still it provides a rescaled tumbling rate that accounts for the fact that crossing events reduce cluster sizes by allowing particles to escape from them.

\section{Mixture equivalence}
\label{equivalence}
Mixtures with different numbers of particle types can be defined as ``equivalent'' if they generate approximately the same CSD. In this context, mapping a mixture to a monodisperse system boils down to the effective monodisperse approach already discussed in Section~\ref{theory}. Proceeding in a similar way, we can now map between two mixtures. As an example, here we briefly discuss the effective monodisperse tumbling rate of a $50$-$50\%$ binary mixture that matches its equivalent fully polydisperse mixture.

This procedure is useful to reduce composition space dimensionality. One example where this would be desirable is as follows. Assume that one wants to study the dynamics of a mixture. In this case, there might exist one equation for the time evolution of each particle type. The total number of equations becomes computationally intractable in the case where many particle types are identified. However, by mapping onto another mixture, we can hope to find a similar system which preserves the dynamical specifics of a mixture---such as non-trivial composition dynamics with multiple evolution stages \cite{PabloPeter1}---while considerably reducing the number of particle types. 

\begin{figure}[!h]
	\centering
	\includegraphics[width=\columnwidth]{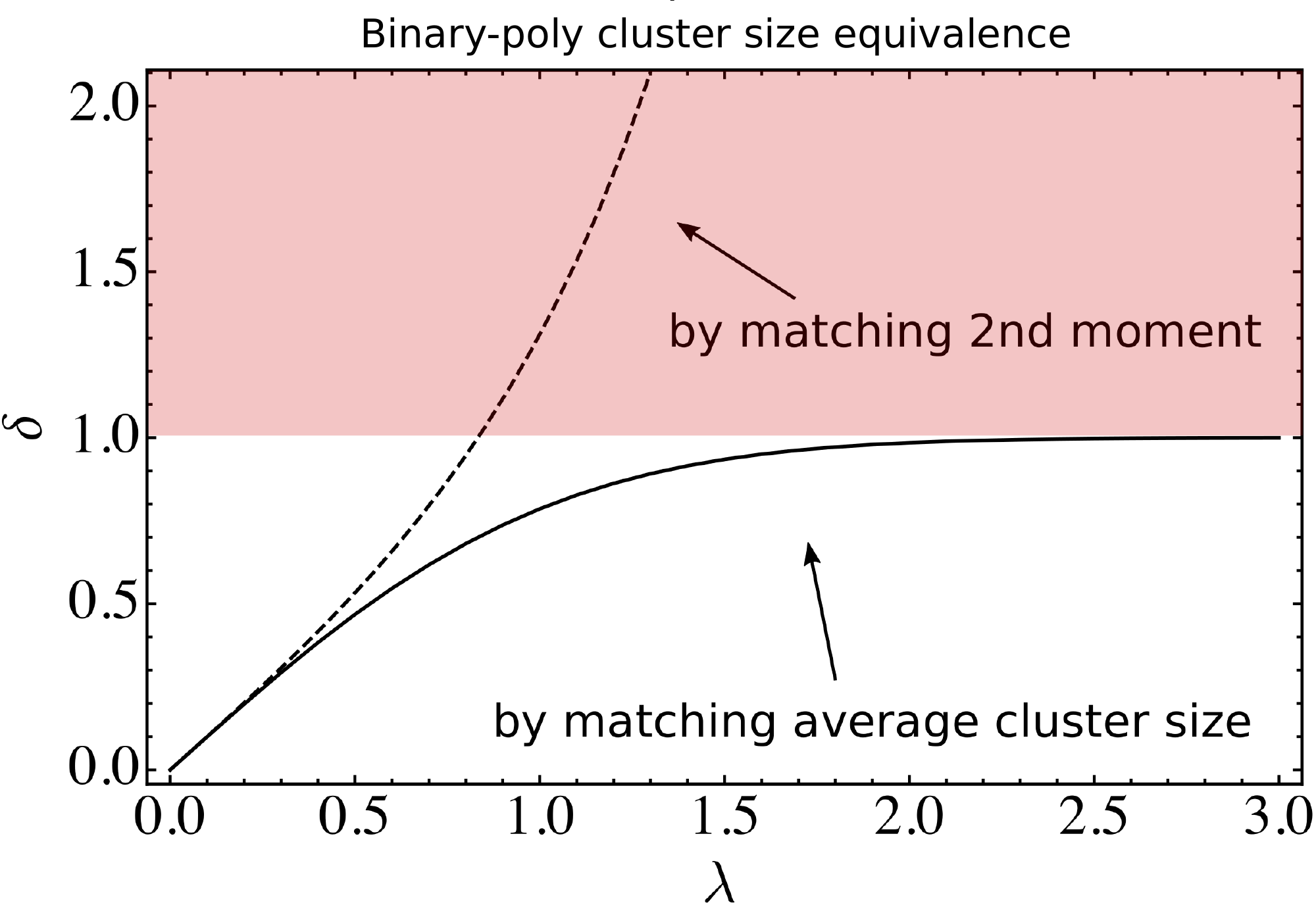}
	\caption{Value of bidispersity $\delta$ which produces a quantitatively similar CSD to a fully polydisperse system with polydispersity distribution parameter $\lambda$. The solid line is obtained by matching the theoretical average cluster sizes between the two mixtures, i.e.\ eqns~\eqref{lengthbi} and \eqref{polylength}. The dashed line comes from matching the first and second statistical moments of the tumbling rate distributions. It enters the region $\delta>1$, which corresponds to unphysical tumbling rates, and then diverges.}
	\label{Fig7}
\end{figure}

By matching the theoretical average cluster sizes between the binary and the fully polydisperse mixtures [eqns~\eqref{lengthbi} and \eqref{polylength}], while fixing that their average tumbling rates are equal, we find a simple relationship between $\delta$ and $\lambda$ as shown in Fig.~\ref{Fig7}, although without closed-form solution.

Another way to map between different mixtures, which has been used in the recent literature, corresponds to matching the tumbling rate distribution moments.\cite{PabloPeter1,PabloPeter2} Considering that a statistical distribution is unambiguously defined by its moments, one proceeds by identifying which parameters of a binary tumbling rate distribution lead to a match of the dominant moments of the fully polydisperse tumbling rate distribution, at least approximately. To second order, we use the mean and the variance, obtaining $\delta=\sqrt{\exp{(\lambda^2)}-1}$. By matching up to the second moment in this way for $\lambda=0.5$, we were able produce virtually the same cluster size distribution between the two simulated mixtures as by matching the average cluster size. The CSDs deviate from one another only at high cluster sizes, as a result of deviations in the higher order moments of the tumbling rate distribution. However, this method fails to match the average cluster sizes as it departs from the first method when one increases $\lambda$, as shown in Fig.~\ref{Fig7}. Furthermore, for $\lambda>\sqrt{\log{2}}$, the matching binary parameter $\delta$ would be $>1$, yielding an unphysical negative value for the lower tumbling rate. Thus a forbidden parameter region for this procedure arises.
On the other hand, the first mapping approach is always possible. In mathematical terms, the first (and correct) method finds the distribution parameters such that both $\langle\alpha\rangle$ and $\langle\alpha^{-1/2}\rangle$ are equal between the binary and fully polydisperse mixtures, while the second method, which fails to predict the average cluster size, looks for the equality of both $\langle\alpha\rangle$ and $\langle\alpha^2\rangle$ between the mixtures.

\section{Conclusions and discussion}
\label{conc}

Motility-induced phase separation is a key phenomenon in active matter. In this work we show how this mechanism leads to clustering amplification induced by motility diversity and clustering suppression induced by confinement relaxation. We used a minimal quasi-1D discrete lattice model of run-and-tumble particles with a distribution of tumbling rates. Neighbouring particles facing each other cross at a constant rate, mimicking a narrow channel whose width is large enough to allow for some position crossing events. A binary mixture and a fully polydisperse system whose distribution is based on experiments were studied. Each particle type contributes to the average cluster size with its own length scale as determined by its tumbling rate. Lower-tumbling-rate particles trap higher-tumbling-rate ones for longer times and therefore contribute with a larger length scale. The average cluster size increases with the degree of motility diversity. Particle crossing allows particles to escape from the clusters more quickly and therefore counteracts the effects of persistent motion. At high enough crossing rate, clustering then becomes controlled by crossing events and motility diversity becomes irrelevant. Our simulations are in good agreement with a maximum entropy-based theory which relies on an effective monodisperse framework and assumes that the tumbling rates obey $\alpha\ll\phi$, the global density. Expressions are found for how the cluster size distribution, the average cluster size, the effective tumbling rate, the steady-state gas density, and the average fraction of mobile particles depend on motility diversity and crossing rate.

\begin{figure}[!htb]
	\centering
	\includegraphics[width=\columnwidth]{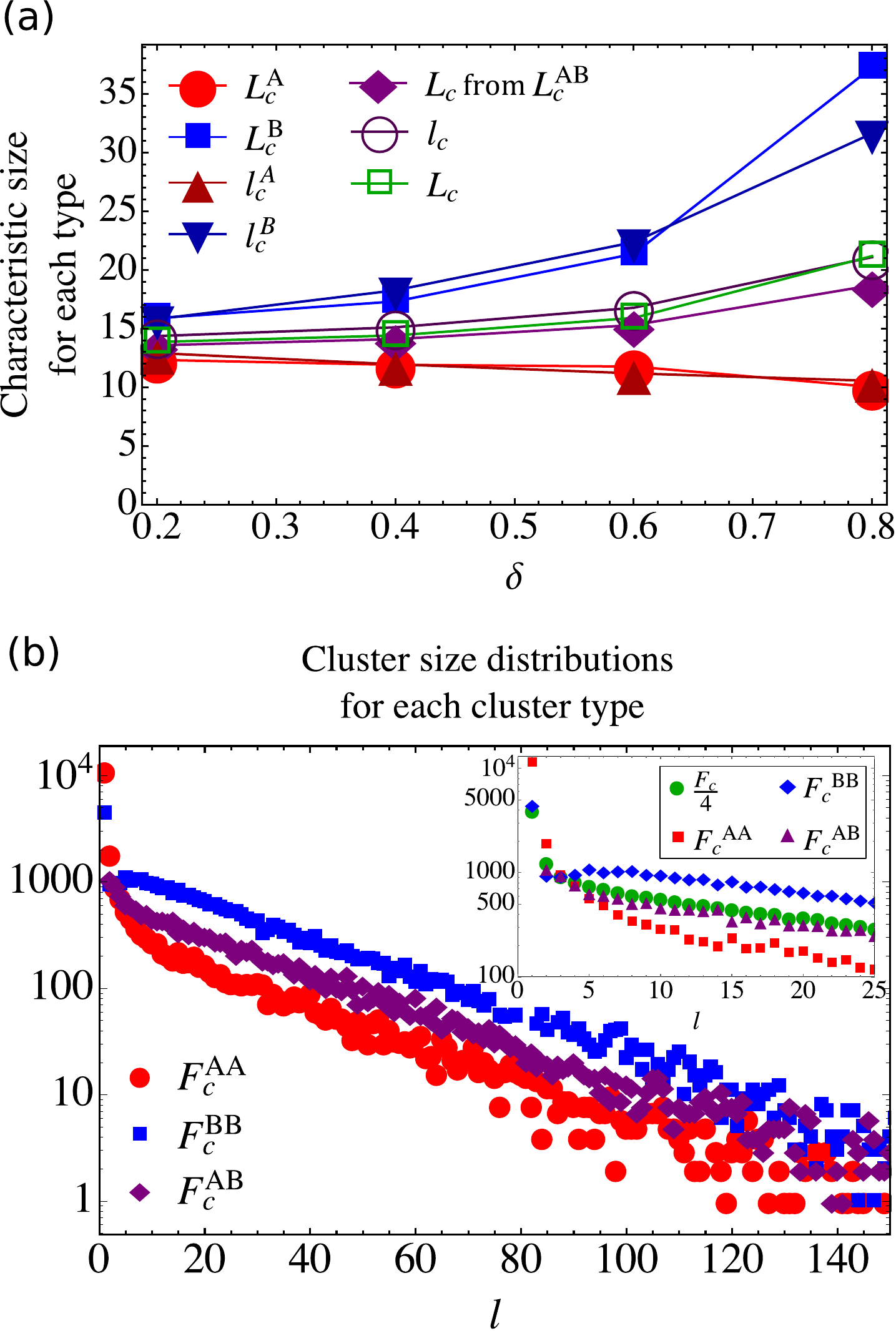}
	\caption{(a) Average cluster sizes ($L_\text{c}^{A}$, $L_\text{c}^{B}$, $L_\text{c}$ from $L_\text{c}^{AB}$, and $L_\text{c}$) and length scales ($l_\text{c}^A$, $l_\text{c}^B$, and $l_\text{c}$) for each particle type and as a function of the bidispersity degree $\delta$ for $\phi=0.5$, $N=2000$, $\alpha_0=0.01$, and $p_{\rm cross}=0$. The former come from averaging partial CSDs where clusters are grouped by border particles types and then multiplying by the appropriate relative fractions (see main text). The latter are the theoretical monodisperse length scales for each particle type and for the system, that is, eqn~\eqref{monoparam} evaluated at the corresponding tumbling rates. (b) Cluster size distributions for each cluster type as defined by the particle types at their borders, i.e.\ $AA$, $AB$, and $BB$ (the case $BA$ is identical to that of $AB$ within numerical error, as expected, and is thus omitted). The bidispersity is $\delta=0.8$. Other parameters as in (a). The inset shows a zoom-in of the small cluster size region. Additionally, it shows $F_c/4$, which agrees with $F_c^{AB}$ (the agreement occurs in the whole range of cluster sizes; data not shown). Here we used $2000$ uncorrelated configurations, within the same simulation, obtained every $10^4$ time steps from ${t=10^7}$ onward. To make this clear, we have \textit{not} (as was done in Fig.~\ref{Fig3}a) normalized the vertical axis by the number of configurations used.}
	\label{Fig9}
\end{figure}

Our theoretical analysis of the simulation data in Section~\ref{theory} was based on the assumption that the average cluster size in a mixture can be predicted from the \emph{average} of the average cluster sizes that one would obtain for each individual particle types. To understand this in more detail we consider again a binary mixture but now we group clusters according to the types of the particles that they have at each end, giving four \emph{cluster types}, i.e.\ $AA$, $AB$, $BA$ and $BB$. We then calculate average cluster sizes for each cluster type and multiply by the relative fractions $f_{AA}/(1/4)$, $f_{AB}/(1/4)$, $f_{BA}/(1/4)$, and $f_{BB}/(1/4)$, where $f_x$ denotes the total number of clusters of type $x$ divided by the total number of clusters. Fig.~\ref{Fig9}a shows that for the $AA$ and $BB$ cluster types the resulting average cluster sizes closely match those theoretically predicted for single particle type systems. This supports our assumption that an understanding of the overall system behaviour can be built up from predictions for those single particle type systems. 

Another way to see this is as follows. For ${\delta=0}$ we have $f_x=1/4$ for all cluster types as there is no physical difference between them. For $\delta>0$, if $f_x$ is bigger (smaller) than $1/4$ for a certain cluster type $x$, then the corresponding average cluster size $l^{x}_\text{c}$ has to be smaller (bigger). This is related to the fact that the cluster sizes are controlled by the border particles and, as just discussed and numerically verified, that ${l_\text{c}^A=4l^{AA}_\text{c} f_{AA}}$ and $l_\text{c}^B=4l^{BB}_\text{c} f_{BB}$. We also find numerically that ${l_\text{c}=4l^{AB}_\text{c} f_{AB}}$ as can be seen from Fig.~\ref{Fig9}a and from the agreement between $F_c/4$ and the distribution of sizes for clusters of type $AB$ in the inset of Fig.~\ref{Fig9}b. Therefore,
\begin{align}
	l_\text{c}&\equiv l^{AA}_\text{c} f_{AA}+l^{BB}_\text{c} f_{BB}+2l^{AB}_\text{c} f_{AB}\nonumber \\ &=\frac{l^A_\text{c}}{4}+\frac{l^B_\text{c}}{4}+\frac{l_\text{c}}{2}
\end{align}
which leads to our assumption used throughout Section~\ref{theory} that ${l_\text{c}=(l^A_\text{c}+l^B_\text{c})/2}$.

Nonetheless, the situation is richer than the above argument would suggest: for example, the size distributions for the individual cluster types are \emph{not} single exponentials (see Fig.~\ref{Fig9}b), except for the $BB$ clusters, i.e.\ the ones bounded on both sides by the more persistent particles. The other distributions show a rapid initial decay followed by an exponential tail with approximately the same exponential decay length. Nevertheless, the overall CSD is close to a single exponential (Fig.~\ref{Fig3}a). How this simple overall behaviour emerges from that of the individual cluster types will be an interesting question for future research.

\begin{figure}[!h]
	\centering
	\includegraphics[width=\columnwidth]{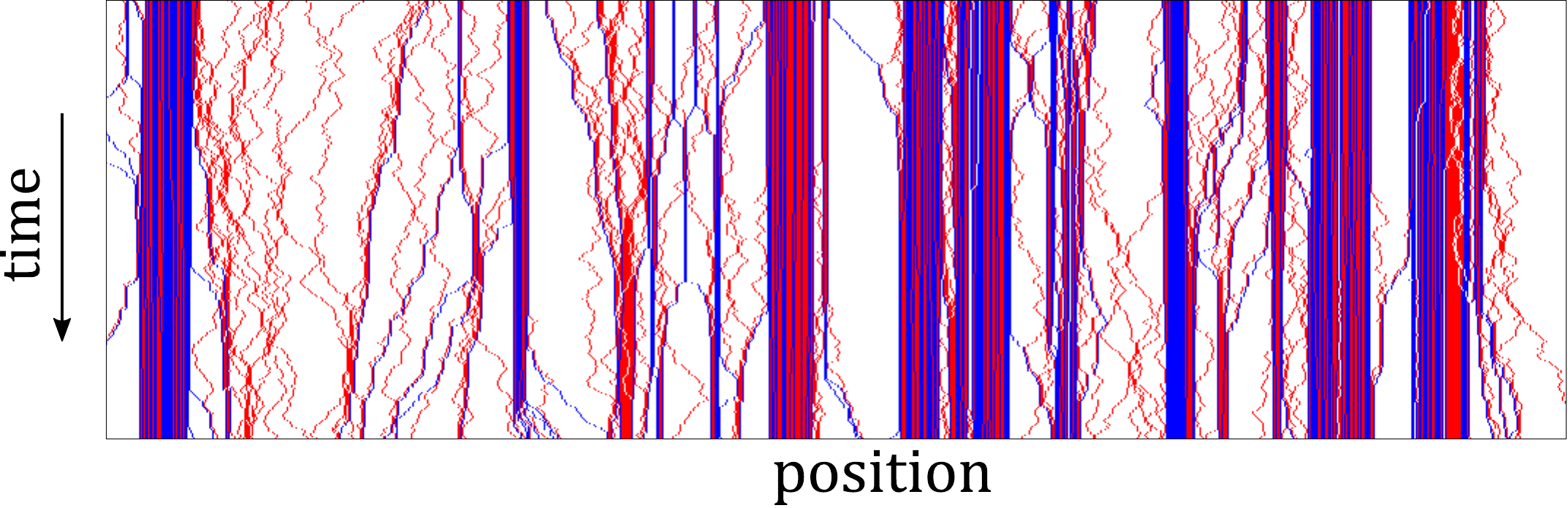}
	\caption{As Fig.~\ref{Fig2} but for $\delta=0.99$,  $\alpha_0=0.2$, $\phi=0.4$, and $p_{\rm cross}=0.01$. Only $1000$ of $N=2000$ simulated sites are shown. As before, particles in red (blue) have higher (lower) tumbling rate.}
	\label{Fig8}
\end{figure}

Particle conservation implies that the global composition, i.e.\ the fractions of particle types, must remain unchanged at all times. If the cluster and gas phases have compositions that are different from the original homogeneous phase (and from each other), then we say that the system has undergone \emph{fractionation}.\cite{warren1999phase,PabloPeter1,PabloPeter3} For small gas concentration $\phi_g$, the cluster phase contains almost all particles, and so it must have roughly the same composition as the original homogeneous phase. In this case, no significant fractionation can occur. At higher $\phi_g$, particles fractionate more significantly. This is especially true when particle crossing is on, allowing particle types to demix across the cluster and gas phases. A thorough quantitative analysis to identify parameter regions where fractionation is maximal is beyond our scope here. Nevertheless, we have been able to find configurations where the gas phase is clearly dominated by higher-tumbling-rate particles in the expected parameter ranges, such as in Fig.~\ref{Fig8}.

One important question is how the average cluster size should depend on the polydisperse distribution parameters in other 1D models of active matter. To answer this, we first notice that, for off-lattice models, it has already been shown through simulations that monodisperse systems of run-and-tumble bacteria, active Brownian particles, and active Ornstein-Uhlenbeck particles, all have an average cluster size of $L_\text{c}\sim\sqrt{\rho u/\omega}$, where $\rho$ is the off-lattice model global concentration, $u$ is the active speed, and $\omega$ is an inverse persistence time parameter.\cite{D0SM00687D} This formula is exactly the off-lattice version of the length scale expression \eqref{monoparam}. Consequently, this strongly indicates that the dependence on the polydisperse distribution parameters is universal in 1D, provided that the polydisperse motility attribute is $\omega$ (or $\alpha$ in our notation). If the polydispersity is in the active speed, which in our lattice model is fixed to $1$ for all particle types, then, following the same procedure as in Section~\ref{theory} for a binary mixture, we would have $L_\text{c}\sim\sqrt{1+\delta }+\sqrt{1-\delta}$, which \emph{decreases} with the diversity degree. The effect of particle crossing in this case is not clear as the frequency of crossing events will strongly depend on the swimming speeds, although moving to an off-lattice framework seems more appropriate to study this.

In future work, several additional features might be included. For instance, depending on the bacteria strain and species, the propulsion mechanism may be so strong that the particles might be able to collectively move and merge clusters. Also, in real systems particle-wall hydrodynamic interactions might substantially change the cluster escape time. Another avenue is to consider effects of finite rather than infinite memory in the evolution of the protein concentration that triggers the tumbles. Finally, one could investigate if there are analogies with passive systems where the polydispersity degree alone can significantly change the phase behaviour of a system. For instance, in the case of passive size-disperse spheres, as the spread of diameters increases, an increasing number of coexisting crystalline equilibrium phases appear.\cite{sollich2010crystalline} That is, by increasing polydispersity, the system enters further into the coexistence region of the phase diagram, moving away from parameter regions where no phase separation occurs. In our active case, something broadly similar happens, in the sense that, by increasing polydispersity in the tumbling rates, the clusters increase in size, therefore moving the system away from parameter regions where the clusters are small, i.e.\ where almost no phase separation occurs.

\section*{Conflicts of interest}
There are no conflicts to declare.

\section*{Acknowledgements}
This research is supported by Fondecyt Grant No.~1180791 (R.S.) and by the Millennium Nucleus Physics of Active Mater of ANID (Chile). We are also thankful to Francisco M.~Rocha for stimulating discussions.



\balance

\renewcommand\refname{References}

\bibliography{Active.bib} 

\providecommand*{\mcitethebibliography}{\thebibliography}
\csname @ifundefined\endcsname{endmcitethebibliography}
{\let\endmcitethebibliography\endthebibliography}{}
\begin{mcitethebibliography}{45}
\providecommand*{\natexlab}[1]{#1}
\providecommand*{\mciteSetBstSublistMode}[1]{}
\providecommand*{\mciteSetBstMaxWidthForm}[2]{}
\providecommand*{\mciteBstWouldAddEndPuncttrue}
  {\def\EndOfBibitem{\unskip.}}
\providecommand*{\mciteBstWouldAddEndPunctfalse}
  {\let\EndOfBibitem\relax}
\providecommand*{\mciteSetBstMidEndSepPunct}[3]{}
\providecommand*{\mciteSetBstSublistLabelBeginEnd}[3]{}
\providecommand*{\EndOfBibitem}{}
\mciteSetBstSublistMode{f}
\mciteSetBstMaxWidthForm{subitem}
{(\emph{\alph{mcitesubitemcount}})}
\mciteSetBstSublistLabelBeginEnd{\mcitemaxwidthsubitemform\space}
{\relax}{\relax}

\bibitem[Cates and Tailleur(2015)]{cates2015motility}
M.~E. Cates and J.~Tailleur, \emph{Annu. Rev. Condens. Matter Phys.}, 2015,
  \textbf{6}, 219--244\relax
\mciteBstWouldAddEndPuncttrue
\mciteSetBstMidEndSepPunct{\mcitedefaultmidpunct}
{\mcitedefaultendpunct}{\mcitedefaultseppunct}\relax
\EndOfBibitem
\bibitem[Soto and Golestanian(2014)]{soto2014run}
R.~Soto and R.~Golestanian, \emph{Physical Review E}, 2014, \textbf{89},
  012706\relax
\mciteBstWouldAddEndPuncttrue
\mciteSetBstMidEndSepPunct{\mcitedefaultmidpunct}
{\mcitedefaultendpunct}{\mcitedefaultseppunct}\relax
\EndOfBibitem
\bibitem[Sep{\'u}lveda and Soto(2016)]{sepulveda2016coarsening}
N.~Sep{\'u}lveda and R.~Soto, \emph{Physical Review E}, 2016, \textbf{94},
  022603\relax
\mciteBstWouldAddEndPuncttrue
\mciteSetBstMidEndSepPunct{\mcitedefaultmidpunct}
{\mcitedefaultendpunct}{\mcitedefaultseppunct}\relax
\EndOfBibitem
\bibitem[Ginot \emph{et~al.}(2018)Ginot, Theurkauff, Detcheverry, Ybert, and
  Cottin-Bizonne]{ginot2018aggregation}
F.~Ginot, I.~Theurkauff, F.~Detcheverry, C.~Ybert and C.~Cottin-Bizonne,
  \emph{Nature Communications}, 2018, \textbf{9}, 1--9\relax
\mciteBstWouldAddEndPuncttrue
\mciteSetBstMidEndSepPunct{\mcitedefaultmidpunct}
{\mcitedefaultendpunct}{\mcitedefaultseppunct}\relax
\EndOfBibitem
\bibitem[Redner \emph{et~al.}(2016)Redner, Wagner, Baskaran, and
  Hagan]{redner2016classical}
G.~S. Redner, C.~G. Wagner, A.~Baskaran and M.~F. Hagan, \emph{Physical Review
  Letters}, 2016, \textbf{117}, 148002\relax
\mciteBstWouldAddEndPuncttrue
\mciteSetBstMidEndSepPunct{\mcitedefaultmidpunct}
{\mcitedefaultendpunct}{\mcitedefaultseppunct}\relax
\EndOfBibitem
\bibitem[Grobas \emph{et~al.}(2020)Grobas, Polin, and
  Asally]{grobas2020swarming}
I.~Grobas, M.~Polin and M.~Asally, \emph{bioRxiv}, 2020\relax
\mciteBstWouldAddEndPuncttrue
\mciteSetBstMidEndSepPunct{\mcitedefaultmidpunct}
{\mcitedefaultendpunct}{\mcitedefaultseppunct}\relax
\EndOfBibitem
\bibitem[Peruani \emph{et~al.}(2012)Peruani, Starru{\ss}, Jakovljevic,
  S{\o}gaard-Andersen, Deutsch, and B{\"a}r]{peruani2012collective}
F.~Peruani, J.~Starru{\ss}, V.~Jakovljevic, L.~S{\o}gaard-Andersen, A.~Deutsch
  and M.~B{\"a}r, \emph{Physical Review Letters}, 2012, \textbf{108},
  098102\relax
\mciteBstWouldAddEndPuncttrue
\mciteSetBstMidEndSepPunct{\mcitedefaultmidpunct}
{\mcitedefaultendpunct}{\mcitedefaultseppunct}\relax
\EndOfBibitem
\bibitem[Buttinoni \emph{et~al.}(2013)Buttinoni, Bialk{\'e}, K{\"u}mmel,
  L{\"o}wen, Bechinger, and Speck]{buttinoni2013dynamical}
I.~Buttinoni, J.~Bialk{\'e}, F.~K{\"u}mmel, H.~L{\"o}wen, C.~Bechinger and
  T.~Speck, \emph{Physical Review Letters}, 2013, \textbf{110}, 238301\relax
\mciteBstWouldAddEndPuncttrue
\mciteSetBstMidEndSepPunct{\mcitedefaultmidpunct}
{\mcitedefaultendpunct}{\mcitedefaultseppunct}\relax
\EndOfBibitem
\bibitem[Levis and Berthier(2014)]{levis2014clustering}
D.~Levis and L.~Berthier, \emph{Physical Review E}, 2014, \textbf{89},
  062301\relax
\mciteBstWouldAddEndPuncttrue
\mciteSetBstMidEndSepPunct{\mcitedefaultmidpunct}
{\mcitedefaultendpunct}{\mcitedefaultseppunct}\relax
\EndOfBibitem
\bibitem[Dandekar \emph{et~al.}(2020)Dandekar, Chakraborty, and
  Rajesh]{dandekar2020hard}
R.~Dandekar, S.~Chakraborty and R.~Rajesh, \emph{arXiv preprint
  arXiv:2006.05980}, 2020\relax
\mciteBstWouldAddEndPuncttrue
\mciteSetBstMidEndSepPunct{\mcitedefaultmidpunct}
{\mcitedefaultendpunct}{\mcitedefaultseppunct}\relax
\EndOfBibitem
\bibitem[Vanhille~Campos \emph{et~al.}(2019)Vanhille~Campos,
  Alarc{\'o}n~Oseguera, Pagonabarraga, Brito, and
  Valeriani]{vanhille2019collective}
C.~Vanhille~Campos, F.~Alarc{\'o}n~Oseguera, I.~Pagonabarraga, R.~Brito and
  C.~Valeriani, \emph{arXiv}, 2019,  arXiv--1912\relax
\mciteBstWouldAddEndPuncttrue
\mciteSetBstMidEndSepPunct{\mcitedefaultmidpunct}
{\mcitedefaultendpunct}{\mcitedefaultseppunct}\relax
\EndOfBibitem
\bibitem[Caprini and Marconi(2020)]{caprini2020time}
L.~Caprini and U.~M.~B. Marconi, \emph{arXiv preprint arXiv:2006.09551},
  2020\relax
\mciteBstWouldAddEndPuncttrue
\mciteSetBstMidEndSepPunct{\mcitedefaultmidpunct}
{\mcitedefaultendpunct}{\mcitedefaultseppunct}\relax
\EndOfBibitem
\bibitem[Dolai \emph{et~al.}(2020)Dolai, Das, Kundu, Dasgupta, Dhar, and
  Kumar]{D0SM00687D}
P.~Dolai, A.~Das, A.~Kundu, C.~Dasgupta, A.~Dhar and K.~V. Kumar, \emph{Soft
  Matter}, 2020, \textbf{16}, 7077--7087\relax
\mciteBstWouldAddEndPuncttrue
\mciteSetBstMidEndSepPunct{\mcitedefaultmidpunct}
{\mcitedefaultendpunct}{\mcitedefaultseppunct}\relax
\EndOfBibitem
\bibitem[Slowman \emph{et~al.}(2016)Slowman, Evans, and
  Blythe]{slowman2016jamming}
A.~Slowman, M.~Evans and R.~Blythe, \emph{Physical review letters}, 2016,
  \textbf{116}, 218101\relax
\mciteBstWouldAddEndPuncttrue
\mciteSetBstMidEndSepPunct{\mcitedefaultmidpunct}
{\mcitedefaultendpunct}{\mcitedefaultseppunct}\relax
\EndOfBibitem
\bibitem[Illien \emph{et~al.}(2020)Illien, de~Blois, Liu, van~der Linden, and
  Dauchot]{illien2020speed}
P.~Illien, C.~de~Blois, Y.~Liu, M.~N. van~der Linden and O.~Dauchot,
  \emph{Physical Review E}, 2020, \textbf{101}, 040602\relax
\mciteBstWouldAddEndPuncttrue
\mciteSetBstMidEndSepPunct{\mcitedefaultmidpunct}
{\mcitedefaultendpunct}{\mcitedefaultseppunct}\relax
\EndOfBibitem
\bibitem[Barberis and Peruani(2019)]{barberis2019phase}
L.~Barberis and F.~Peruani, \emph{The Journal of Chemical Physics}, 2019,
  \textbf{150}, 144905\relax
\mciteBstWouldAddEndPuncttrue
\mciteSetBstMidEndSepPunct{\mcitedefaultmidpunct}
{\mcitedefaultendpunct}{\mcitedefaultseppunct}\relax
\EndOfBibitem
\bibitem[Whitman \emph{et~al.}(1998)Whitman, Coleman, and
  Wiebe]{whitman1998prokaryotes}
W.~B. Whitman, D.~C. Coleman and W.~J. Wiebe, \emph{Proceedings of the National
  Academy of Sciences}, 1998, \textbf{95}, 6578--6583\relax
\mciteBstWouldAddEndPuncttrue
\mciteSetBstMidEndSepPunct{\mcitedefaultmidpunct}
{\mcitedefaultendpunct}{\mcitedefaultseppunct}\relax
\EndOfBibitem
\bibitem[Tu and Grinstein(2005)]{tu2005white}
Y.~Tu and G.~Grinstein, \emph{Physical Review Letters}, 2005, \textbf{94},
  208101\relax
\mciteBstWouldAddEndPuncttrue
\mciteSetBstMidEndSepPunct{\mcitedefaultmidpunct}
{\mcitedefaultendpunct}{\mcitedefaultseppunct}\relax
\EndOfBibitem
\bibitem[Villa-Torrealba \emph{et~al.}(2020)Villa-Torrealba, Ch\'avez-Raby,
  de~Castro, and Soto]{andrea2020}
A.~Villa-Torrealba, C.~Ch\'avez-Raby, P.~de~Castro and R.~Soto, \emph{Phys.
  Rev. E}, 2020, \textbf{101}, 062607\relax
\mciteBstWouldAddEndPuncttrue
\mciteSetBstMidEndSepPunct{\mcitedefaultmidpunct}
{\mcitedefaultendpunct}{\mcitedefaultseppunct}\relax
\EndOfBibitem
\bibitem[de~Castro and Sollich(2017)]{PabloPeter1}
P.~de~Castro and P.~Sollich, \emph{Phys. Chem. Chem. Phys.}, 2017, \textbf{19},
  22509--22527\relax
\mciteBstWouldAddEndPuncttrue
\mciteSetBstMidEndSepPunct{\mcitedefaultmidpunct}
{\mcitedefaultendpunct}{\mcitedefaultseppunct}\relax
\EndOfBibitem
\bibitem[Warren(1999)]{warren1999phase}
P.~B. Warren, \emph{Physical Chemistry Chemical Physics}, 1999, \textbf{1},
  2197--2202\relax
\mciteBstWouldAddEndPuncttrue
\mciteSetBstMidEndSepPunct{\mcitedefaultmidpunct}
{\mcitedefaultendpunct}{\mcitedefaultseppunct}\relax
\EndOfBibitem
\bibitem[de~Castro and Sollich(2018)]{PabloPeter2}
P.~de~Castro and P.~Sollich, \emph{The Journal of Chemical Physics}, 2018,
  \textbf{149}, 204902\relax
\mciteBstWouldAddEndPuncttrue
\mciteSetBstMidEndSepPunct{\mcitedefaultmidpunct}
{\mcitedefaultendpunct}{\mcitedefaultseppunct}\relax
\EndOfBibitem
\bibitem[de~Castro and Sollich(2019)]{PabloPeter3}
P.~de~Castro and P.~Sollich, \emph{Soft Matter}, 2019, \textbf{15},
  9287--9299\relax
\mciteBstWouldAddEndPuncttrue
\mciteSetBstMidEndSepPunct{\mcitedefaultmidpunct}
{\mcitedefaultendpunct}{\mcitedefaultseppunct}\relax
\EndOfBibitem
\bibitem[de~Castro~Melo(2019)]{decastro2019}
P.~S. de~Castro~Melo, \emph{Phase separation of polydisperse fluids}, King's
  College London, 2019\relax
\mciteBstWouldAddEndPuncttrue
\mciteSetBstMidEndSepPunct{\mcitedefaultmidpunct}
{\mcitedefaultendpunct}{\mcitedefaultseppunct}\relax
\EndOfBibitem
\bibitem[Stenhammar \emph{et~al.}(2015)Stenhammar, Wittkowski, Marenduzzo, and
  Cates]{stenhammar2015activity}
J.~Stenhammar, R.~Wittkowski, D.~Marenduzzo and M.~E. Cates, \emph{Physical
  Review Letters}, 2015, \textbf{114}, 018301\relax
\mciteBstWouldAddEndPuncttrue
\mciteSetBstMidEndSepPunct{\mcitedefaultmidpunct}
{\mcitedefaultendpunct}{\mcitedefaultseppunct}\relax
\EndOfBibitem
\bibitem[Kolb and Klotsa(2020)]{kolb2020active}
T.~Kolb and D.~Klotsa, \emph{Soft Matter}, 2020, \textbf{16}, 1967--1978\relax
\mciteBstWouldAddEndPuncttrue
\mciteSetBstMidEndSepPunct{\mcitedefaultmidpunct}
{\mcitedefaultendpunct}{\mcitedefaultseppunct}\relax
\EndOfBibitem
\bibitem[Hoell \emph{et~al.}(2019)Hoell, L{\"o}wen, and Menzel]{hoell2019multi}
C.~Hoell, H.~L{\"o}wen and A.~M. Menzel, \emph{The Journal of Chemical
  Physics}, 2019, \textbf{151}, 064902\relax
\mciteBstWouldAddEndPuncttrue
\mciteSetBstMidEndSepPunct{\mcitedefaultmidpunct}
{\mcitedefaultendpunct}{\mcitedefaultseppunct}\relax
\EndOfBibitem
\bibitem[Wittkowski \emph{et~al.}(2017)Wittkowski, Stenhammar, and
  Cates]{wittkowski2017nonequilibrium}
R.~Wittkowski, J.~Stenhammar and M.~E. Cates, \emph{New Journal of Physics},
  2017, \textbf{19}, 105003\relax
\mciteBstWouldAddEndPuncttrue
\mciteSetBstMidEndSepPunct{\mcitedefaultmidpunct}
{\mcitedefaultendpunct}{\mcitedefaultseppunct}\relax
\EndOfBibitem
\bibitem[Takatori and Brady(2015)]{takatori2015theory}
S.~C. Takatori and J.~F. Brady, \emph{Soft Matter}, 2015, \textbf{11},
  7920--7931\relax
\mciteBstWouldAddEndPuncttrue
\mciteSetBstMidEndSepPunct{\mcitedefaultmidpunct}
{\mcitedefaultendpunct}{\mcitedefaultseppunct}\relax
\EndOfBibitem
\bibitem[Grosberg and Joanny(2015)]{grosberg2015nonequilibrium}
A.~Grosberg and J.-F. Joanny, \emph{Physical Review E}, 2015, \textbf{92},
  032118\relax
\mciteBstWouldAddEndPuncttrue
\mciteSetBstMidEndSepPunct{\mcitedefaultmidpunct}
{\mcitedefaultendpunct}{\mcitedefaultseppunct}\relax
\EndOfBibitem
\bibitem[Curatolo \emph{et~al.}(2020)Curatolo, Zhou, Zhao, Liu, Daerr,
  Tailleur, and Huang]{curatolo2020cooperative}
A.~Curatolo, N.~Zhou, Y.~Zhao, C.~Liu, A.~Daerr, J.~Tailleur and J.~Huang,
  \emph{Nature Physics}, 2020,  1--6\relax
\mciteBstWouldAddEndPuncttrue
\mciteSetBstMidEndSepPunct{\mcitedefaultmidpunct}
{\mcitedefaultendpunct}{\mcitedefaultseppunct}\relax
\EndOfBibitem
\bibitem[Wang \emph{et~al.}(2020)Wang, Shen, Xia, Feng, and
  Tian]{wang2020phase}
Y.~Wang, Z.~Shen, Y.~Xia, G.~Feng and W.~Tian, \emph{Chinese Physics B}, 2020,
  \textbf{29}, 053103\relax
\mciteBstWouldAddEndPuncttrue
\mciteSetBstMidEndSepPunct{\mcitedefaultmidpunct}
{\mcitedefaultendpunct}{\mcitedefaultseppunct}\relax
\EndOfBibitem
\bibitem[van~der Meer \emph{et~al.}(2020)van~der Meer, Prymidis, Dijkstra, and
  Filion]{van2020predicting}
B.~van~der Meer, V.~Prymidis, M.~Dijkstra and L.~Filion, \emph{The Journal of
  Chemical Physics}, 2020, \textbf{152}, 144901\relax
\mciteBstWouldAddEndPuncttrue
\mciteSetBstMidEndSepPunct{\mcitedefaultmidpunct}
{\mcitedefaultendpunct}{\mcitedefaultseppunct}\relax
\EndOfBibitem
\bibitem[Quelas \emph{et~al.}(2016)Quelas, Althabegoiti, Jimenez-Sanchez,
  Melgarejo, Marconi, Mongiardini, Trejo, Mengucci, Ortega-Calvo, and
  Lodeiro]{quelas2016swimming}
J.~I. Quelas, M.~J. Althabegoiti, C.~Jimenez-Sanchez, A.~A. Melgarejo, V.~I.
  Marconi, E.~J. Mongiardini, S.~A. Trejo, F.~Mengucci, J.-J. Ortega-Calvo and
  A.~R. Lodeiro, \emph{Scientific Reports}, 2016, \textbf{6}, 23841\relax
\mciteBstWouldAddEndPuncttrue
\mciteSetBstMidEndSepPunct{\mcitedefaultmidpunct}
{\mcitedefaultendpunct}{\mcitedefaultseppunct}\relax
\EndOfBibitem
\bibitem[Ranjard and Richaume(2001)]{ranjard2001quantitative}
L.~Ranjard and A.~Richaume, \emph{Research in microbiology}, 2001,
  \textbf{152}, 707--716\relax
\mciteBstWouldAddEndPuncttrue
\mciteSetBstMidEndSepPunct{\mcitedefaultmidpunct}
{\mcitedefaultendpunct}{\mcitedefaultseppunct}\relax
\EndOfBibitem
\bibitem[M{\"a}nnik \emph{et~al.}(2009)M{\"a}nnik, Driessen, Galajda, Keymer,
  and Dekker]{mannik2009bacterial}
J.~M{\"a}nnik, R.~Driessen, P.~Galajda, J.~E. Keymer and C.~Dekker,
  \emph{Proceedings of the National Academy of Sciences}, 2009, \textbf{106},
  14861--14866\relax
\mciteBstWouldAddEndPuncttrue
\mciteSetBstMidEndSepPunct{\mcitedefaultmidpunct}
{\mcitedefaultendpunct}{\mcitedefaultseppunct}\relax
\EndOfBibitem
\bibitem[Fuentes-Ramirez and Caballero-Mellado(2005)]{fuentes2005bacterial}
L.~E. Fuentes-Ramirez and J.~Caballero-Mellado, in \emph{PGPR: Biocontrol and
  biofertilization}, Springer, 2005, pp. 143--172\relax
\mciteBstWouldAddEndPuncttrue
\mciteSetBstMidEndSepPunct{\mcitedefaultmidpunct}
{\mcitedefaultendpunct}{\mcitedefaultseppunct}\relax
\EndOfBibitem
\bibitem[Ribeiro and Potiguar(2016)]{ribeiro2016active}
H.~Ribeiro and F.~Potiguar, \emph{Physica A: Statistical Mechanics and its
  Applications}, 2016, \textbf{462}, 1294--1300\relax
\mciteBstWouldAddEndPuncttrue
\mciteSetBstMidEndSepPunct{\mcitedefaultmidpunct}
{\mcitedefaultendpunct}{\mcitedefaultseppunct}\relax
\EndOfBibitem
\bibitem[Vicsek \emph{et~al.}(1995)Vicsek, Czir{\'o}k, Ben-Jacob, Cohen, and
  Shochet]{vicsek1995novel}
T.~Vicsek, A.~Czir{\'o}k, E.~Ben-Jacob, I.~Cohen and O.~Shochet, \emph{Physical
  Review Letters}, 1995, \textbf{75}, 1226\relax
\mciteBstWouldAddEndPuncttrue
\mciteSetBstMidEndSepPunct{\mcitedefaultmidpunct}
{\mcitedefaultendpunct}{\mcitedefaultseppunct}\relax
\EndOfBibitem
\bibitem[Chen and Berg(2000)]{chen2000torque}
X.~Chen and H.~C. Berg, \emph{Biophysical Journal}, 2000, \textbf{78},
  1036--1041\relax
\mciteBstWouldAddEndPuncttrue
\mciteSetBstMidEndSepPunct{\mcitedefaultmidpunct}
{\mcitedefaultendpunct}{\mcitedefaultseppunct}\relax
\EndOfBibitem
\bibitem[Cluzel \emph{et~al.}(2000)Cluzel, Surette, and
  Leibler]{cluzel2000ultrasensitive}
P.~Cluzel, M.~Surette and S.~Leibler, \emph{Science}, 2000, \textbf{287},
  1652--1655\relax
\mciteBstWouldAddEndPuncttrue
\mciteSetBstMidEndSepPunct{\mcitedefaultmidpunct}
{\mcitedefaultendpunct}{\mcitedefaultseppunct}\relax
\EndOfBibitem
\bibitem[Figueroa-Morales \emph{et~al.}(2020)Figueroa-Morales, Soto, Junot,
  Darnige, Douarche, Martinez, Lindner, and Cl{\'e}ment]{figueroa20183d}
N.~Figueroa-Morales, R.~Soto, G.~Junot, T.~Darnige, C.~Douarche, V.~A.
  Martinez, A.~Lindner and {\'E}.~Cl{\'e}ment, \emph{Physical Review X}, 2020,
  \textbf{10}, 021004\relax
\mciteBstWouldAddEndPuncttrue
\mciteSetBstMidEndSepPunct{\mcitedefaultmidpunct}
{\mcitedefaultendpunct}{\mcitedefaultseppunct}\relax
\EndOfBibitem
\bibitem[Mandal \emph{et~al.}(2020)Mandal, Bhuyan, Chaudhuri, Dasgupta, and
  Rao]{mandal2020extreme}
R.~Mandal, P.~J. Bhuyan, P.~Chaudhuri, C.~Dasgupta and M.~Rao, \emph{Nature
  Communications}, 2020, \textbf{11}, 1--8\relax
\mciteBstWouldAddEndPuncttrue
\mciteSetBstMidEndSepPunct{\mcitedefaultmidpunct}
{\mcitedefaultendpunct}{\mcitedefaultseppunct}\relax
\EndOfBibitem
\bibitem[Lovely and Dahlquist(1975)]{lovely1975statistical}
P.~S. Lovely and F.~Dahlquist, \emph{Journal of Theoretical Biology}, 1975,
  \textbf{50}, 477--496\relax
\mciteBstWouldAddEndPuncttrue
\mciteSetBstMidEndSepPunct{\mcitedefaultmidpunct}
{\mcitedefaultendpunct}{\mcitedefaultseppunct}\relax
\EndOfBibitem
\bibitem[Sollich and Wilding(2010)]{sollich2010crystalline}
P.~Sollich and N.~B. Wilding, \emph{Physical Review Letters}, 2010,
  \textbf{104}, 118302\relax
\mciteBstWouldAddEndPuncttrue
\mciteSetBstMidEndSepPunct{\mcitedefaultmidpunct}
{\mcitedefaultendpunct}{\mcitedefaultseppunct}\relax
\EndOfBibitem
\end{mcitethebibliography}
\bibliographystyle{rsc} 

\end{document}